\documentclass[10pt,journal,compsoc]{IEEEtran}
\ifCLASSOPTIONcompsoc
  \usepackage[nocompress]{cite}
\else
  \usepackage{cite}
\fi
\usepackage{bbm}
\usepackage{url}
\usepackage{stfloats}
\usepackage{amsfonts}
\usepackage[dvips]{graphicx}
\usepackage{cite}
\usepackage{amsmath}
\usepackage{array}
\usepackage{amssymb}

\usepackage{stfloats}
\usepackage{graphicx}
\usepackage{footnote}
\usepackage{booktabs}
\usepackage{array}
\usepackage{algorithm}
\usepackage{subeqnarray}
\usepackage{cases}
\usepackage{threeparttable}
\usepackage{color}
\usepackage{textcomp}

\usepackage{epstopdf}
\usepackage{algpseudocode}
\usepackage{bm}
\usepackage{multirow}
\usepackage[labelformat=simple]{subcaption}
\usepackage{adjustbox}

\newtheorem{proposition}{Proposition}
\newtheorem{assumption}{Assumption}
\newtheorem{lemma}{Lemma}

\usepackage{cite}
\usepackage{graphicx}

\ifCLASSINFOpdf
\else
\fi
\begin{document}
%
\title{Average-Case Analysis of Greedy Matching for Large-Scale D2D Resource Sharing}
%
%
%
%

\author{Shuqin~Gao,
        Costas~A.~Courcoubetis
        and~Lingjie~Duan,~\IEEEmembership{Senior~member,~IEEE}
         \thanks{The work of Lingjie Duan was supported by the Ministry of Education, Singapore, under its Academic Research Fund Tier 2 Grant under Award MOE-T2EP20121-0001.}

}

\IEEEtitleabstractindextext{%
\begin{abstract}

Given the proximity of many wireless users and their diversity in consuming local resources (e.g., data-plans, computation and energy resources), device-to-device (D2D) resource sharing is a promising approach towards realizing a sharing economy. This paper adopts an easy-to-implement greedy matching algorithm with distributed fashion and only sub-linear $O(\log n)$ parallel complexity (in user number $n$) for large-scale D2D sharing. Practical cases indicate that the greedy matching's average performance is far better than the worst-case approximation ratio $50\%$ as compared to the optimum. However, there is no rigorous average-case analysis in the literature to back up such encouraging findings and this paper is the first to present such analysis for multiple representative classes of graphs. For 1D linear networks, we prove that our greedy algorithm performs better than $86.5\%$ of the optimum. For 2D grids, though dynamic programming cannot be directly applied, we still prove this average performance ratio to be above $76\%$. For the more challenging Erdos-R{\'e}nyi random graphs, we equivalently reduce to the asymptotic analysis of random trees and successfully prove a ratio up to $79\%$. Finally, we conduct experiments using real data to simulate realistic D2D networks, and show that our analytical performance measure approximates well practical cases. 
\end{abstract}

}

\maketitle

\IEEEdisplaynontitleabstractindextext

%
\IEEEpeerreviewmaketitle

\IEEEraisesectionheading{\section{Introduction}\label{sec:introduction}}

%
%
%
%
\IEEEPARstart{T}{hanks} to advances in wireless and smartphone technologies, mobile users in proximity can use local wireless links (e.g., short-range communications) to share local resources (e.g., data-plans \cite{xuehe,data}, computation \cite{d2dfogging,D2Dcomputing}, caching memory \cite{caching,caching3} and energy \cite{energy,p2ppower}). For instance, in a busy airport, subscribed users who have leftover data plans can set up personal/portable hotspots and share data connections to travelers with high roaming fees \cite{xuehe}; in a crowded stadium, users with unutilized storage can download faster and share the cached popular game videos with other users in the vicinity \cite{caching}; or in an exposition, users who would like to watch the product introductory videos can use cooperative video streaming to share video segments with each other \cite{microcast}. Given the large diversity for each user in the levels of her individual resource utilization, device-to-device (D2D) resource sharing is envisioned as a promising approach to pool resources and increase social welfare.


Some recent studies have been conducted for modeling and guiding D2D resource sharing in wireless networks (e.g., [2-16]). As a node in the established D2D network graph, each mobile user can be a resource consumer or supplier, depending on whether her local resource is sufficient or not. As in \cite{D2Dcomputing} and \cite{caching}, according to their locations, each user can only connect to a subset of users in the neighborhood through wireless connections, and the available wireless links are modelled as edges in the network graph. Sharing between any two connected users brings in certain benefit to the pair, which is modelled as a non-negative weight to the corresponding edge.

All these works optimize resource allocation by matching users in a centralized manner that requires global information and strict coordination. Hence the developed approaches cannot scale well in a scenario involving a large number of users, due to a large communication and computation overhead caused by the centralized nature of the proposed solutions. Carrying this argument further, the existing optimal weighted matching algorithms from the literature cannot be effectively used in the case of large user-defined networks due to their centralized nature and super-linear time complexity \cite{overview}. This motivates the need for developing distributed algorithms that exploit parallelism, have low computation complexity and good average performance for practical parameter distributions.

In the broader literature of distributed algorithm design for matching many nodes in a large graph, a greedy matching algorithm of linear complexity is proposed in \cite{paper7} and \cite{paper8} without requiring a central controller. It simply selects each time the edges with local maximum weights and yields an approximation ratio of $1/2$ as compared to the optimum. A parallel algorithm is further proposed in \cite{logtime} to reduce complexity at the cost of obtaining a smaller approximation ratio than $1/2$. It should be noted that in the analysis of these algorithms, complexity and approximation ratio are always worst-case measures, but the worst-case approximation ratio rarely happens in most network cases in practice. This work is motivated by our observation from the simulation that the greedy matching’s average performance is far better than the worst-case approximation ratio of $50\%$ as compared to the optimum, being at least $95\%$ of the optimum in most cases. To our best knowledge, this work is the first analytical study to present an average-case performance analysis of distributed matching algorithms. The results of our average-case analysis are important in practice because they motivate the use of such simple greedy matching algorithms without substantial performance degradation.

Since worst-case bounds no longer work for average-case analysis, we develop totally new techniques to analyze average performance. These techniques become more accurate when taking into account the structure of the network graph, and provide a very positive assessment of the greedy matching's average performance that is far from the worst case. Since the greedy matching can be naturally implemented in parallel by each node in the network, we also prove that with high probability (w.h.p.), the algorithm has sub-linear parallel complexity $O(\log n)$ in the number of users $n$. Our main contributions are summarized as follows. 

\begin{itemize}

\item \emph{Average-case analysis of greedy matching in large-scale regular networks:} For large-scale 1D linear networks, we first use a new graph decomposition method to compute the upper bound for the optimal matching and then derive a recursive formula for the greedy matching by using dynamic programming. We prove that our greedy algorithm performs at least better than $86.5\%$ of the optimum, and the minimum ratio is achieved when all the edges take similar weight values. In 2D grids, the same analysis cannot be directly applied. We introduce a new asymptotic analysis method based on truncating the 2D grids and then manage to analyze the resulting specific sub-grids using a recursive calculation similar to the 1D case. We prove that our greedy matching's average performance ratio is still above $76\%$. For these types of graphs, our greedy algorithm has only sub-linear complexity $O(\log n)$ w.h.p.. Thus, our algorithm provides a great implementation advantage compared to the optimal matching algorithms that require super-linear complexity without sacrificing much on performance.

\item \emph{Average-case analysis of greedy matching in large-scale random graphs:} Besides the grids of fixed topology, we develop a new theoretic technique to analyze large Erdos-R{\'e}nyi random graphs $G(n, p)$, where each of $n$ users connects to any other user with probability $p$. For a dense random graph with constant $p$, we prove that the greedy matching will almost surely provide the highest possible total matching value, leading to an average performance ratio that tends to $100\%$ as $n$ increases. The analysis of sparse graphs with $p<1/n$ is more challenging, but we reduce it to the asymptotic analysis of random trees since the probability of the existence of loops in the sparse random graphs is zero w.h.p.. By exploiting the recursive nature of trees, we derive a recursive formula for the greedy matching, which is not closed-form but can be solved using bisection. Finally, we manage to obtain rigorous average performance bounds and parallel complexity $O(\log n)$ w.h.p.. The average performance ratio reaches its minimum (still above $79\%$) when the graph is neither dense nor sparse.
    
    \item \emph{Extension to multi-unit resource sharing:} We extend from single-unit to multi-unit resource sharing in our model, where each user may have multiple units of local resources to share. Our greedy algorithm in the multi-unit version requires parallel complexity $O(\log n)$ w.h.p.. By developing a new graph decomposition method, we prove that its average performance ratio is at least $78\%$.

    \item \emph{Application to practical scenarios:} We conduct experiments using real data for mobile user locations to simulate realistic D2D networks with constraints on the maximum allowed communication distance between devices. We show that our analytical $G(n,p)$ performance measure approximates well practical cases of such D2D sharing networks. To decide the maximum D2D sharing range among users, we take into account the D2D communication failure due to path-loss and mutual interference among matched pairs. The optimal sharing range is achieved by finding the best tradeoff between transacting with more devices but at the higher risk that the chosen best neighbor might not be effectively usable due to a communication failure.


\end{itemize}

The paper is organized as follows. In Section \ref{sec:related}, we discuss the related work and emphasize how it differs from our work. In Section \ref{sec:preliminaries}, we present our network model and the greedy matching algorithm for solving the D2D resource sharing problem in any network graph. In Sections \ref{sec:linear} and \ref{sec:grid}, we analyze the average performance ratios of the algorithm in the 1D and 2D grids. Sections \ref{sec:random} and \ref{sec:multiple} extend the average-case analysis to random graphs and multi-unit resource sharing. Section \ref{sec:practical} shows simulation results for application to practical scenarios and Section \ref{sec:conclusions} concludes the paper.

\section{Related Work}\label{sec:related}

We discuss the related work concerning the two main topics related to our paper, namely D2D resource sharing ideas and distributed matching algorithms. 

Recent research efforts have been devoted to D2D resource sharing due to advances in wireless and smartphone technologies. In \cite{xuehe} and \cite{data}, subscribed users who have leftover data plans can set up personal/portable hotspots and share data connections with those who face data deficits. In \cite{d2dfogging} and \cite{D2Dcomputing}, mobile users can share the computation resources with each other by task offloading via cellular D2D links. In \cite{opportunistic1} and \cite{opportunistic2}, mobile devices can collaborate with each other to process and deliver data over D2D channels to fulfill crowdsourcing tasks. In \cite{caching2}, unmanned aerial vehicles (UAVs) with residual cache capacity can help store contents for others using inter-UAV connections. \cite{momd} and \cite{md} allow mobile video users to support others in proximity to download video segments through WiFi or Bluetooth. However, compared to our work, the key focus in these D2D resource sharing works is to design a way to match the supply with the demand locally, not to analyze performances theoretically.

In the theoretical literature on the classical maximum matching problem, most existing methods to solve it require a central controller to gather all participants’ information and perform the computation centrally \cite{overview}. This severely hinders the scalability of large-scale D2D sharing. A distributed matching algorithm using the primal-dual method is proposed in \cite{c3} to find the optimum, but requires a prohibitively high average computational complexity. There are some recent works focusing on finding approximation distributed algorithms that run fast \cite{survey}. In particular, two log-time parallel algorithms (with respect to user number) are proposed in \cite{logtime} and \cite{logtime2} for general graphs, and another faster algorithm is proposed to compute an efficient matching in an expected constant time for the special case of tree graphs in \cite{tree}. But these algorithms' complexity and approximation ratio are only analyzed in the worst case and may not hold in most cases. Our work is the first analytical study to present an average-case performance analysis of distributed matching algorithms.


\section{System Model and Problem Formulation}\label{sec:preliminaries}

\subsection{System Model for D2D Resource Sharing}

We first describe our D2D resource sharing model that involves a large number of potential users to share resources with each other via local wireless links (e.g., short-range communications). In this model, resources are exchanged between participating users in repeated rounds. In each round, we first run an algorithm to determine how to match users that have sent requests to neighbors for exchanging resources in this round, and then realize the actual sharing of the corresponding resources as determined by the algorithm. The set of participating users and the available D2D links may be different for different rounds. 

In each round, we depict the network graph as $G=(U,E)$, where $U$ is the set of nodes corresponding to users that are participating in the given round, and $E$ is the set of D2D links between participating users that are feasible to establish with some minimum level of performance (e.g., signal strength, actual physical distance, etc., depending on the application). Our model reasonably assumes a time-scale separation between the time for users to change location and the time to perform a round of the algorithm (that typically should be in the order of a few seconds), in order to establish stable D2D communication for resource sharing [2-16]. This might not hold in the case of fast-moving cars but it is the case when users typically hang around in crowded urban areas with low mobility, e.g., walking streets, airports, stadiums, city parks, cafes, malls, etc.. For each user $u_i\in U=\{u_1,u_2,\dots,u_n\}$, the subset $A(u_i)\subseteq U$ denotes the set of her neighbors in $G$, i.e., there is a feasible D2D link $e_{ij}\in E$ between $u_i$ and any $u_j\in A(u_i)$. Note that different definitions of `feasibility' for D2D links will imply a different set of edges $E$ between the users in $U$. Also the set $U$ is changing over time/rounds since new users may join the sharing economy and existing users may drop out after satisfying their needs or moving out of range.

With each edge $e_{ij}\in E$, there is an associated weight $w_{ij}\geq 0$ that models the surplus (or welfare gain) of sharing a unit resource between users $i$ and $j$ if they are `matched' in our terminology, usually converted in some monetary basis (say $\$$). Let $W=\{w_{ij}\}$ be the weight vector over all edges of $G$. Note that our model is very flexible and can fit various applications of D2D resource sharing by allowing for different ways to define the values for $w_{ij}$. In the case of a consumer $i$ with revenue $r_i$ for obtaining a unit resource and a supplier $j$ with cost $c_j$ for offering a unit resource, the weight is clearly $w_{ij}= r_i-c_j$. For example, in a secondary data-plan trading market \cite{xuehe,data}, user $u_i$ with data-plan surplus shares her personal hotspot connection with neighboring user $u_j$ with high roaming fee, and weight $w_{ij}$ models the difference between user $u_j$'s saved roaming fee and the sharing cost (e.g., energy consumption in battery) of user $u_i$. In another example of cooperative video streaming \cite{momd}, user $u_i$ seeks user $u_j$'s assistance to download and share video segments via a local wireless connection so as to improve the streaming experience. The quality of experience (QoE), which is usually referred as to user perception, is measured in terms of download time or video rate in this example. Then, $w_{ij}$ becomes the difference between the QoE improvement (in some appropriate units) of user $u_i$ and the download/sharing cost of user $u_j$. 

Besides the cases where nodes are partitioned into suppliers and consumers, there are certain applications where edges capture the effects of collaboration between users if these users are matched. A simple example is the exchange of information where both parties benefit (e.g., \cite{caching,microcast}). 
Suppose that users $i$, $j$ cache the sets $F_i$ and $F_j$ of popular files respectively,
and assume that each user has files that the other user would also like to have. 
If they get matched by the algorithm, they will exchange a total $|F_i\cup F_j|-|F_i\cap F_j|$ files, and the total social benefit $w_{ij}$ can be approximated to be proportional to the above number
or some more accurate estimate of the value of the shared information. We have constructed a case study of such collaborative caching in a network graph based on some real data in Section \ref{sec:caching}.

\begin{figure}[t]\centering
\vspace{-0cm}
\includegraphics[width=6.3cm]{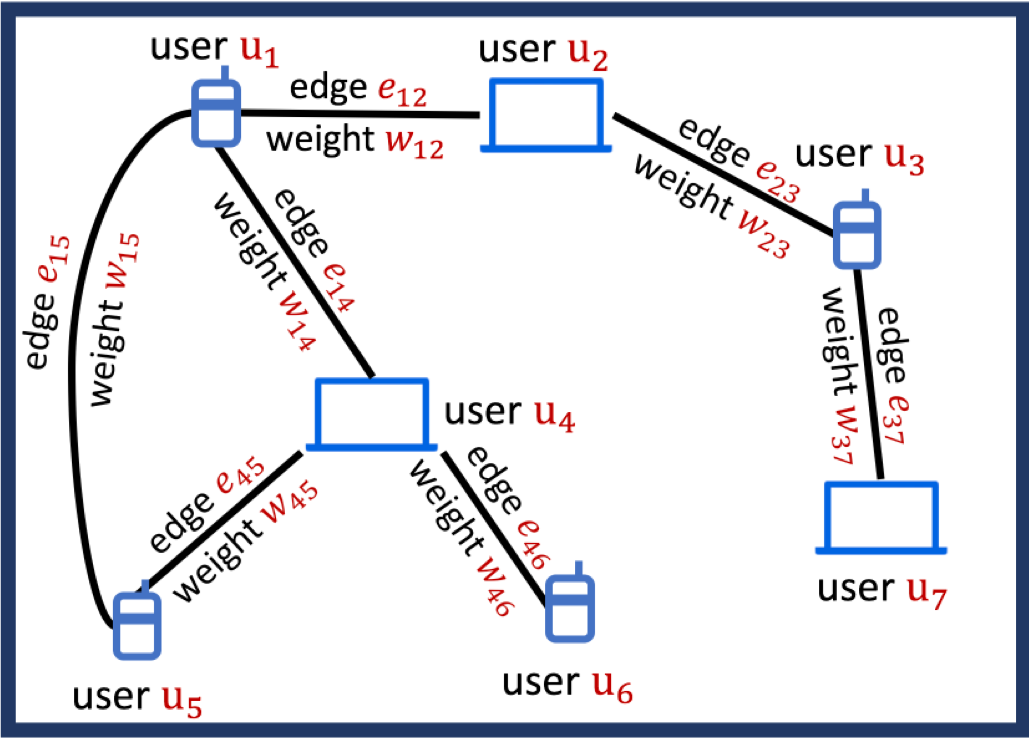}
\captionsetup{font=footnotesize}
\caption{An illustrative instance of the D2D resource sharing model with $n=7$ users is captured spatially.}
\label{fig:instance}
\vspace{-0cm}
\end{figure}

In any given round, our sharing model corresponds to an instance of a random weighted graph $(G=(U,E),W)$. A simple interpretation of the model is that a typical user, when participating, corresponds to a randomly selected node in $G$. In particular, we don't care for the actual identity of the participating users (after all, we care for the total value generated in the economy, summed over all participants). To simplify the model, we assume certain properties for the resulting stochastic process, i.e., in each round the set $U$ and the corresponding $E$, $W$ are independent identically distributed (IID), with certain distributions. In particular, we assume that the weights $w_{ij}$ take values from a finite discrete set $V\hspace{-2pt}=\hspace{-2pt}\{v_1,v_2,\dots,v_{K}\}$ according to the general probability distribution $Pr(w_{ij}\hspace{-2pt}=\hspace{-2pt}v_k)\hspace{-2pt}=p_k$ with $\sum_{k=1}^K p_k=1$. Without loss of generality, we assume $0\hspace{-2pt}\leq \hspace{-2pt}v_1\hspace{-2pt}<\hspace{-2pt}v_2\hspace{-2pt}<\hspace{-2pt}\cdots\hspace{-2pt}<\hspace{-2pt}v_{K}$. A small-scale illustrative instance of the D2D resource sharing model is shown spatially on the ground in Fig.~\ref{fig:instance}, which can be abstracted to a weight graph $(G\hspace{-2pt}=\hspace{-2pt}(U\hspace{-2pt}=\hspace{-2pt}\{u_1,u_2,\dots,u_7\},E\hspace{-2pt}=\hspace{-2pt}\{e_{12},e_{14},e_{15},e_{23},e_{37},e_{45},e_{46},\}),W\hspace{-2pt}=\hspace{-2pt}\{w_{12},w_{14},w_{15},w_{23},w_{37},w_{45}, w_{46}\})$.

In typical practices of D2D sharing (e.g., energy transfer), a user is only matched to a single neighbor (if any) to finally transact with\footnote{Allowing more concurrent matchings per user might not greatly improve performance, since our simulations suggest that most of the total benefit is usually obtained from one among the possible matchings where the values of the matchings follow a Pareto distribution.}. Keeping this simple but practical case of single matching per user\footnote{We extend to multi-unit resource sharing with similar results in Section \ref{sec:multiple}.}, given a weighted graph $(G=(U,E),W)$, we would like to select the pairs of users to match in order to maximize the total sharing benefit (i.e., the `social welfare'). Assuming full and globally available information on $G$ and $W$, we formulate the social welfare maximization problem as a maximum weighted matching problem:
\begin{subequations}
\begin{align}
&\mathcal{P}_1:&\max \quad & \sum_{e_{ij}\in E} w_{ij}x_{ij}, \label{equ:objective}\\
&&\text {s.t.} \quad & \sum_{u_j\in A(u_i)} x_{ij}\leq 1, \quad \forall u_i\in U, \label{equ:matching1}\\
&&& x_{ij}\in \{0,1\},\quad\forall e_{ij}\in E, \label{equ:assignment}
\end{align}
\end{subequations}
where $x_{ij}$ is the binary optimization variable denoting whether edge $e_{ij}$ is included in the final matching ($x_{ij}=1$) or not ($x_{ij}=0$). Constraint \eqref{equ:matching1} tells that any user $u_i$ can only be matched to at most one user in her set of neighbors $A(u_i)$.

\subsection{Preliminaries of Greedy Algorithm}

According to \cite{overview}, to optimally solve the maximum weighted matching problem $\mathcal{P}_1$, one needs to centrally gather the weight and graph connectivity information beforehand. Further, searching for all possible matchings results in super-linear computation complexity, which is formidably high for a large-scale network with a large number $n$ of users. Alternatively, the greedy matching addresses these two issues by keeping information local and allowing the algorithm to be performed in a distributed fashion. Algorithm 1 outlines the key steps of the greedy matching algorithm (please see intuition in the text that follows).

\vspace{0.1cm}
\noindent\textbf{Algorithm 1:} Greedy matching algorithm for solving problem $\mathcal{P}_1$ for the graph ($G=(U,E), W$).

\textbf{Initialization:} $U'=U$; $A'(u_i) = A(u_i), \forall u_i\in U $; $x_{ij}$

$=0,\forall e_{ij}\in E$.

In each iteration, repeat the following two phases:

\textbf{Proposal phase:} 

For each unmatched user $u_i\in U'$:

\begin{itemize}
    \item User $u_i$ selects a user
    $u_{j^*}$ among her unmatched neighbors in $A'(u_i)$ with the maximum weight $w_{ij^*}$.
    \item User $u_i$ sends to $u_{j^*}$ a matching proposal.
\end{itemize}

\textbf{Matching phase:} 

For a user pair $(u_i,u_j)$ that both $u_i$ and $u_j$ receive 

proposals from each other:

\begin{itemize}
    \item Match $u_i$ and $u_j$ by updating $x_{ij}=1$ and $U'=U'\setminus \{u_i,u_j\}$.
    \item Make $u_i$ and $u_j$ unavailable for matching with others, by updating $A'(u_{k})=A'(u_{k})\setminus \{u_i\}$ for any $u_{k}\in A'(u_i)$, and similarly for $u_j$.

\end{itemize}

\begin{figure}[t]\centering
\vspace{-0cm}
\includegraphics[width=5.5cm]{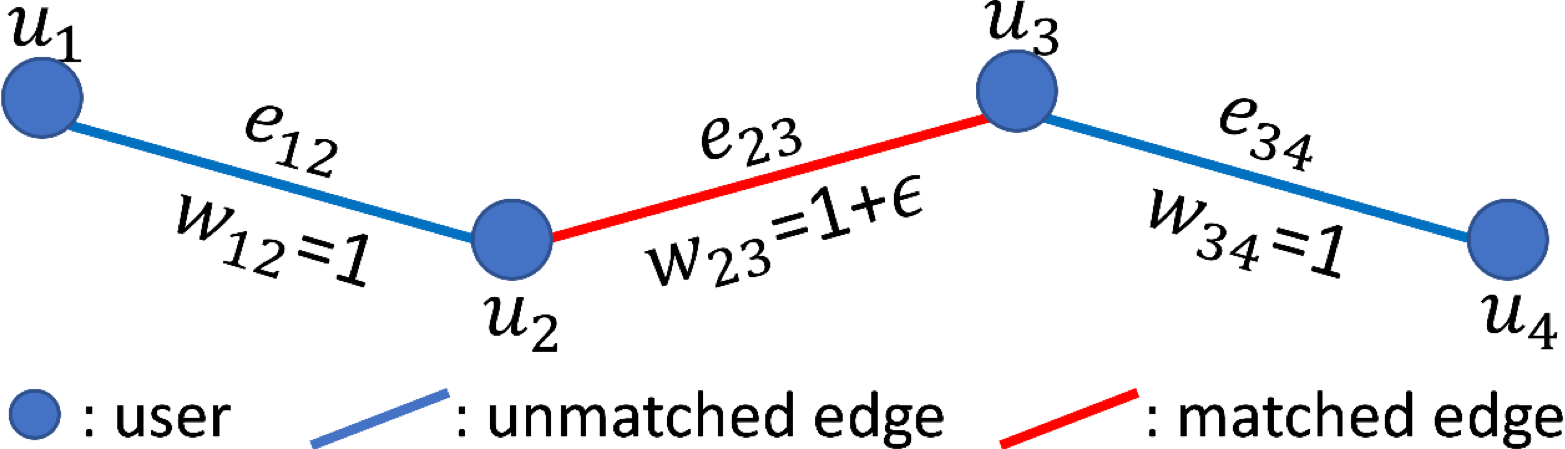}
\captionsetup{font=footnotesize}
\caption{A simple example of approximation ratio of $1/2$ achieved by Algorithm 1. The greedy matching returned by Algorithm 1 is the red-colored edge $e_{23}$ with weight $1+\epsilon, \epsilon>0$, while the optimal matching is $\{e_{12},e_{34}\}$ in blue with total weight $2$. The corresponding approximation ratio is $(1+\epsilon)/2$, taking its minimum value when $\epsilon\rightarrow 0^+$.}
\label{fig:systemmodel}
\vspace{-0.3cm}
\end{figure}

First note that Algorithm 1 is randomized in the selection of preferred neighbors in case there are multiple equally best choices in the proposal phase. A way to simplify this and make the algorithm deterministic is to assume that nodes are assigned unique numbers and that a node assigns priority in the case of ties to its neighbor with the highest number. This avoids loops and guarantees termination in $O(|E|)$ steps. In the rest of the paper, we can assume this deterministic version for Algorithm 1. This is a mild assumption that shall not affect the validity of our key analysis. More importantly, Algorithm 1 can be implemented \emph{distributedly}: at each time, each user uses local information to choose the unmatched neighbor with the highest weight as her potential matching partner; she will stop once this preference becomes reciprocal, or there are no available unmatched neighbors. This algorithm calculates a matching with total weight at least $1/2$ of the optimum (see \cite{paper8}). This worst-case approximation ratio of $1/2$ is achieved in the instance in Fig.~\ref{fig:systemmodel} when $\epsilon\to 0^+$, since the greedy matching chooses the middle edge while the optimal matching chooses the two side edges. Besides, when considering the instability of the connections between matched pairs (e.g., due to users' mobility or network failure), we prove that if a fraction $a\%$ of devices become disconnected in the middle of the matching transaction, the social welfare is reduced by less than $2a\%$ on average due to the different sharing alternatives available to the remaining nodes, and failures not being correlated with the value of the matching that would take place. This is another robustness property of Algorithm 1.

\subsection{Our Problem Statement for Average-Case Analysis}

Although the approximation ratio of Algorithm 1 is $1/2$ with half efficiency loss in the worst case, see \cite{paper7}, \cite{paper8}, this ratio is achieved in Fig.~\ref{fig:systemmodel} only when the middle edge has slightly larger weight than its two adjacent edges. In such a simple three-edge instance, given that weights of independent edges are equally likely to be either $1$ or $1+\epsilon$, the worst-case approximation ratio $50\%$ happens only with probability $1/4$. Here, our greedy algorithm still performs better than $87.5\%$ of the optimum in the average sense\footnote{Here, when running our greedy algorithm, we follow that the four nodes $u_1$, $u_2$, $u_3$ and $u_4$ in Fig.~\ref{fig:systemmodel} are assigned decreasing ID values (i.e., decreasing priority over ties among neighbors whose edge has the same weight).}. 

In a large-scale network instance, given the IID distribution of the choice of the weights,
it is more improbable that the graph will consist of an infinite repetition of the above special weighted three-edge pattern which leads to the worst-case performance. 
Hence, we expect the average performance ratio of the greedy matching to be much greater than $1/2$. 

Since worst-case bound no longer works for average-case analysis, we aim to develop totally new techniques to theoretically analyze the average performance of representative classes of graphs with random parameters. To start with, we first provide the rigorous definitions for our average-case performance analysis.

By taking expectation with respect to the weights in $W$ that are IID with a general discrete distribution $Pr(w_{ij}=v_k)=p_k,\forall k=1,\ldots,K$, we define the average performance ratio $PR(G)$ of Algorithm 1 for a given graph $G$ as follows:
\begin{align}
PR(G)=\frac{\mathbb{E}_{W}[\hat{f}(G,W)=\sum_{e_{ij}\in E} w_{ij}\hat{x}_{ij}]}{\mathbb{E}_{W}[f^\star(G,W)=\sum_{e_{ij}\in E} w_{ij} x^{\star}_{ij}]},\label{equ:average}
\end{align}
where $f^\star(G,W)$ and $\hat{f}(G,W)$ denote the total weights (i.e., social welfare) under the optimal matching and the greedy matching, respectively, $\{x^\star_{ij}\}, \{\hat{x}_{ij}\}$ being the corresponding matchings. Since over time the algorithm is repeated for new instances, the numerator and denominator correspond to the time-average of the social welfare obtained by running the greedy and the optimal algorithms, respectively. 

We next evaluate the performance ratio for several special forms of practical interest for $G$ that corroborate the excellent performance of the greedy matching, including the large-scale 1D and 2D grids of fixed topology, as well as the random graph $G(n,p)$ networks. In the case of random graphs, we must take expectation in \eqref{equ:average} over both $G$ and $W$. Besides, we will also prove the sub-linear computation complexity to run Algorithm 1 for these large-scale networks.

\section{Average-Case Analysis for D2D Sharing in 1D Linear Networks}\label{sec:linear}

\begin{figure}[t]\centering
\vspace{-0cm}
\includegraphics[width=8.5cm]{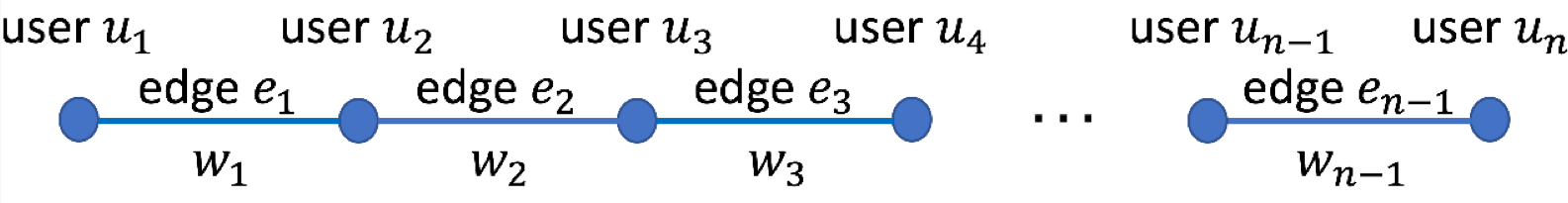}
\captionsetup{font=footnotesize}
\caption{The large-scale linear network model with $n$ users.}
\label{fig:line}
\vspace{-0.3cm}
\end{figure}

When many users are distributed in an avenue or road and can locally share their resources (e.g., walking along 5th Av. at Christmas), we may use a 1D linear network to approximate their connectivity and analyze the greedy matching's average performance. 1D linear networks are the simplest case of regular graphs and are used as a theoretical device to get insights for our later average-case analysis of 2D regular graphs, the more general random graphs and the extension to multi-unit resource case in Sections \ref{sec:grid}, \ref{sec:random} and \ref{sec:multiple}. As illustrated in Fig.~\ref{fig:line}, we consider a large weighted linear network, where each user $u_i$ (except for starting and ending users $u_1$ and $u_n$) locally connects with two adjacent users $u_{i-1}$ and $u_{i+1}$. In such linear networks, for notational simplicity we use $e_i$ instead of $e_{i,i+1}$ to denote the connection between users $u_{i}$ and $u_{i+1}$, and similarly use weight $w_i$ instead of $w_{i, i+1}$. The corresponding weight vector becomes $W=\{w_1,w_2,\dots,w_{n-1}\}$.

For the linear network with $n$ users as shown in Fig.~\ref{fig:line}, we first analyze the running time of Algorithm 1, where a unit of time corresponds to one iteration of the steps of Algorithm 1. We simulate the system in practice by running the greedy matching in parallel by each node. Different from the related literature (e.g., \cite{paper7,paper8}), we focus on analyzing the parallel complexity of Algorithm 1 below. 

Let $H(u_i)$ denote the length of the longest chain (sequence of edges) that has non-decreasing weights and starts from $u_i$ towards the left or right side. Suppose that $w_{i-1}\leq w_{i-2}\leq\cdots\leq w_{i-H(u_i)+1}\leq w_{i-H(u_i)} >w_{i-H(u_i)-1}$ is the longest chain. We claim that $u_i$ will terminate running Algorithm 1 (i.e., by being matched or knowing that it has no available unmatched neighbors) within $H(u_i)/2$ time. This is easy to see since starting from time 0, the edge $e_{i-H(u_i)}$ will be included in the total matching in iteration 1, $e_{i-H(u_i)+2}$ in iteration 2, etc. Hence, in less than $H(u_i)/2$ steps, all neighbors of $u_i$ will have resolved their possible preferences towards users different than $u_i$, and subsequently $u_i$ will either be matched with one of her neighbors or be left with an empty unmatched neighbor set. 

As Algorithm 1 terminates when all users make their final decisions, if the probability of any user in $G$ having a chain longer than $c\log n$ (i.e., $\max_{u_i\in U} H(u_i)> c\log n$) for some constant $c$ is very small, then the parallel execution of Algorithm 1 will terminate within $O(\log n)$ time with very high probability. This is the case for large-scale linear networks as the next proposition states. Note that in the literature, \cite{paper8} just proves linear $O(n)$ time bound for running the greedy matching (but the execution is not parallel).
\begin{proposition}\label{pro:complexitylinear}
In large-scale linear networks of $n$ users, Algorithm 1 runs in $O(\log n)$ time w.h.p.. 
\end{proposition}

The proof is given in Appendix \ref{app:complexitylinear} of the Supplementary Material of this TMC submission. Next, we focus on studying the average performance ratio $PR(G)$ in \eqref{equ:average}. The exact value of the average total weight $\mathbb{E}_{W}[f^\star(G,W)]$ under the optimal matching is difficult to analyze due to formidably many matching combinations over the large network. We aim to derive a lower bound for $PR(G)$, by first deriving an upper bound for the denominator $\mathbb{E}_{W}[f^\star(G,\hspace{-2pt}W)]$ in \eqref{equ:average}, and then obtaining an exact asymptotic expression for the numerator $\mathbb{E}_{W}[\hat{f}(G,W)]$ in \eqref{equ:average}.

\subsection{Average Performance Analysis of Optimal Matching}\label{sec:optimal}

\begin{figure}[t]\centering
\vspace{-0.2cm}
\includegraphics[width=8.5cm]{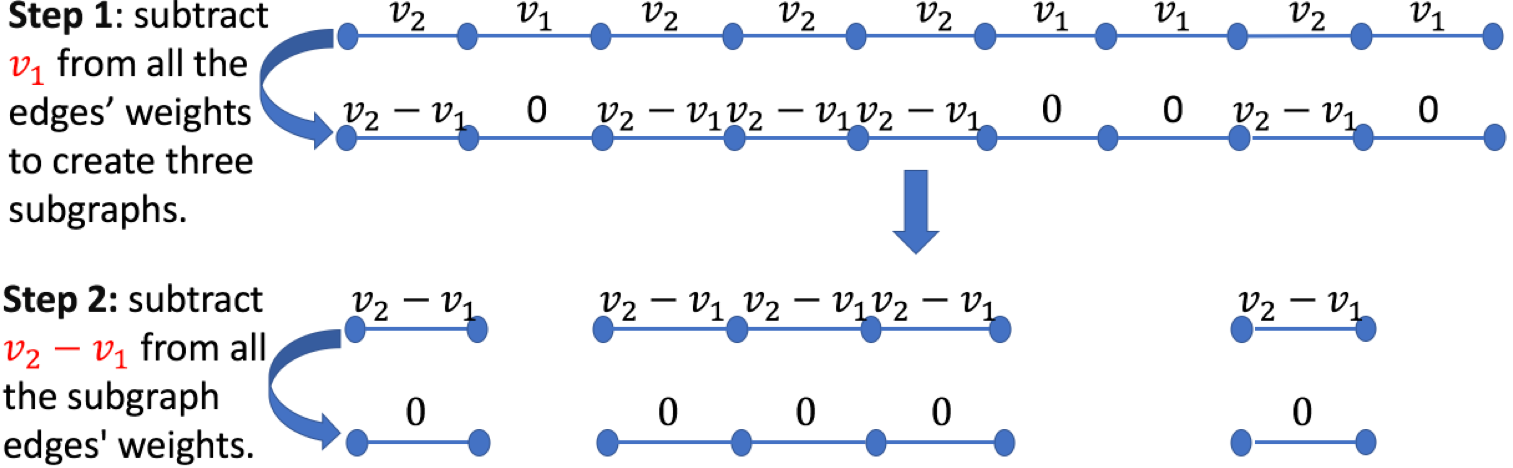}
\captionsetup{font=footnotesize}
\caption{Illustration of the graph decomposition method for a linear network example with $K=2$ possible edge weights $v_1<v_2$. In the first step, we subtract $v_1$ from all the $9$ edges' weights, creating $4$ zero-weight edges that reduce the graph into three linear sub-graphs of size $1$, $3$ and $1$. In the second step, for each sub-graph, we subtract $v_2-v_1$ to result in no edges with positive weights. An upper bound for the optimal matching is $\lceil\frac{9}{2}\rceil v_1+(1+\lceil\frac{3}{2}\rceil+1)(v_2\hspace{-1pt}-v_1)=v_1+4v_2$, obtained in a straightforward fashion using the total subtracted weights from the maximum cardinality matching for each step.}
\label{fig:optimaldeduct}
\vspace{-0.3cm}
\end{figure}

To find the upper bound on the average total weight $\mathbb{E}_{W} [f^\star(G,W)]$ under the optimal matching, we propose a new graph decomposition method to reduce network connectivity, by creating edges with zero weight value purposely. Such edges do not contribute to the total weight of the matching, simplifying the optimal matching of the reduced graph. 

Note that in our weight set $V$, there are $K$ possible weight values satisfying $v_1<v_2<\cdots<v_{K}$. Our method's basic idea is to reduce the original network into a large number of disconnected components, by subtracting from all edge weights first $v_1$, then $v_2-v_1$, $v_3-v_2$, etc. This procedure takes $K$ steps to conclude until creating a graph consisting of zero-weight edges. An illustrative example for
$K=2$ is shown in Fig.~\ref{fig:optimaldeduct}, where we take two steps to obtain the performance upper bound of the optimum. The total amount of weights that are subtracted from the maximum cardinality matching in the reduced graph during each of the $K$ steps is an upper bound for the optimal matching. In the next proposition we analytically obtain the closed-form upper bound for $\mathbb{E}_{W}[f^\star(G,W)]$ for any linear network.
\begin{proposition} \label{pro:upperoptimal}
Given the weight set $V=\{v_1,v_2,\dots,v_K\}$ with the weight distribution $P=\{p_1,p_2,\dots,p_K\}$, the average total weight of the optimal matching in large-scale linear networks of $n$ users is upper bounded by
\begin{align}
\mathbb{E}_{W} \hspace{-2pt}[f^\star(G,W)]\hspace{-2pt}\leq \hspace{-2pt} n\frac{v_1}{2}\hspace{-2pt}+\hspace{-2pt}n\hspace{-2pt}\sum_{k=1}^{K-1} (v_{k+1}\hspace{-2pt}-\hspace{-2pt}v_k)\frac{1\hspace{-2pt}-\hspace{-2pt}\sum_{i=1}^k p_i}{2\hspace{-2pt}-\hspace{-2pt}\sum_{i=1}^k p_i}.\label{equ:optimalupper}
\end{align}
\end{proposition}

The proof is given in Appendix \ref{app:upperoptimal} of the Supplementary Material of this TMC submission. The upper bound in \eqref{equ:optimalupper} is linearly increasing in user number $n$ and increases in weight value $v_k$ for any $k\in\{1,2,\dots,K\}$. This bound is tight only when edges can take a single weight value, i.e., $K=1$.

\subsection{Average Performance Analysis of Algorithm 1}\label{sec:greedy}

Without loss of generality, when running Algorithm 1, we suppose that each user facing the same weights of the two adjacent edges assigns higher priority to match with the left-hand-side neighbor in Fig.~\ref{fig:line}.

\begin{assumption}
For each user $u_i$ having the same weights $w_{i-1}=w_{i}$ with the two adjacent neighbors $u_{i-1}$ and $u_{i+1}$, Algorithm 1 assigns higher priority to match with the left-side neighbor $u_{i-1}$ (see Fig.~\ref{fig:line}).\end{assumption}

This makes Algorithm 1 deterministic and returns a unique solution. We prove the following lemma. 

\begin{lemma}\label{lem:firstk}
Given the weight set size $K$, an edge $e_i$ that satisfies $w_i>w_{i-1}$ and $w_i\geq w_{i+1}$ can be found within the first $K$ edges of the linear network graph.
\end{lemma}

\begin{figure}[t]\centering
\vspace{-0.2cm}
\includegraphics[width=8cm]{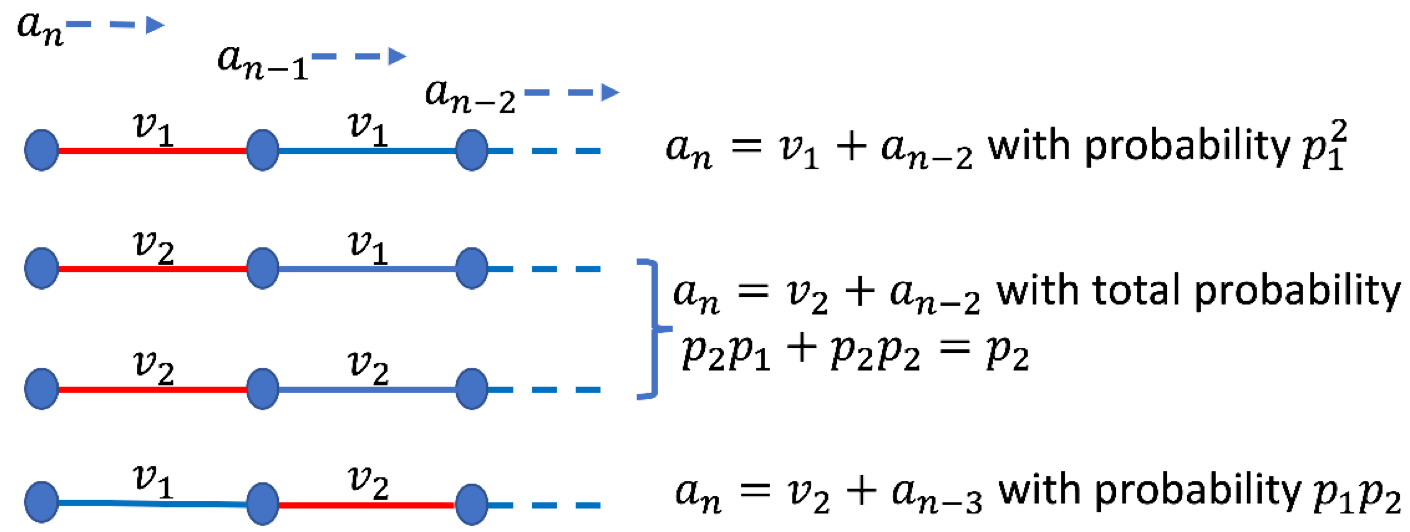}
\captionsetup{font=footnotesize}
\caption{Given the weight set size $K=2$ and $v_1<v_2$, an edge that certainly matches is always found within the first $2$ edges, as marked in red. There are totally $K^K=4$ weight combination cases of the first $2$ edges. In each case, after matching the red edge, the remaining graph is still linear but with a smaller size $n-2$ or $n-3$.}
\label{fig:k2}
\vspace{-0.3cm}
\end{figure}

Fig.~\ref{fig:k2} shows an illustrative example for $K=2$, and we always find such an edge (marked in red) with local maximum weight within the first 2 edges. This edge will be matched in Algorithm 1, and the remaining graph is still linear but with a smaller user size. Then, we reduce the total matching into two sub-problems: the matching of the edges from $e_1$ to $e_i$ and the matching of the remaining edges to the right. Given such reduction, we are able to derive the recursive formula for calculating the result of the greedy matching by using dynamic programming. More specifically, by considering all the $K^K$ weight combinations $\{w_1, \dots, w_K\}$ of the first $K$ edges and the existence of edge $e_i$ that will certainly match, we derive the \emph{recursive formula for the sequence $\{a_n\}$}, where $a_n$ denotes the average total weight of the greedy matching with $n$ users. In the example of $K=2$ in Fig.~\ref{fig:k2}, there are four weight combination cases, where each realized case has a recursive formula. By taking the expectation with respect to the probabilities of the four cases, the expected recursive formula is given by
\begin{align}
&a_n=p_1^2(v_1+a_{n-2})+p_2(v_2+a_{n-2})+p_1p_2(v_2+a_{n-3})\nonumber\\
&=p_1^2v_1+(p_2+p_1p_2)v_2+(p_2+p_1^2)a_{n-2}+p_1p_2a_{n-3}.\label{equ:greedyrecurrence}
\end{align}
Based on this, we derive $a_n=\frac{p_1^2v_1+(p_2+p_1p_2)v_2}{2p_2+2p_1^2+3p_1p_2}n+o(n)$ by using asymptotic analysis. 

Moreover, for an arbitrary $K$, it is also possible to derive the recursive formula for the sequence $\{a_n\}$ as a function of $V=\{v_1,v_2,\dots,v_K\}$ and $P=\{p_1,p_2,\dots,p_K\}$. For a uniform weight distribution (i.e., $p_1\hspace{-2pt} =\hspace{-2pt} p_2\hspace{-2pt} =\hspace{-2pt} \cdots \hspace{-2pt} = \hspace{-2pt} p_K \hspace{-2pt} = \hspace{-2pt} 1/K$), this simplifies and we obtain the following closed-form result.

\begin{proposition}\label{pro:generalanyk}
For an arbitrary $K$, if $p_1=\cdots=p_K=\frac{1}{K}$, the recursive formula for the sequence $\{a_n\}$ in large-scale linear networks is given by
\begin{align}
a_n=\sum_{k=1}^K \beta_k v_k+ \gamma_k a_{n-k-1}, \nonumber
\end{align}
where $\beta_k=\frac{(K-1)^{K-k}}{K^K(K+1)^{-k+1}}$ and $\gamma_k=\frac{1}{K^{k+1}} \sum\limits_{i=k}^K i {{i-1}\choose{k-1}}$. By applying asymptotic analysis for a large user number $n$, we derive closed-form $a_n$ for our greedy matching's performance below:
\begin{align}
a_n=\frac{\sum_{k=1}^K \beta_kv_k}{\sum_{k=1}^K (k+1)\gamma_k}n+o(n). \label{equ:anyk}
\end{align}
\end{proposition}

The proof is given in Appendix \ref{app:generalanyk} of the Supplementary Material of this TMC submission. \eqref{equ:anyk} is useful later in Sections \ref{sec:prlinear} and \ref{sec:grid} to derive the performance guarantee of Algorithm 1 in linear and grid networks.

\subsection{Average Performance Ratio of Algorithm 1}\label{sec:prlinear}

We first check the special case of weight set size $K=2$. Based on \eqref{equ:optimalupper} and the general formula derived by \eqref{equ:greedyrecurrence}, we obtain the closed-form average performance ratio of Algorithm 1 as compared to the optimal matching.
\begin{proposition}\label{pro:guarantee2}
In large-scale linear networks with $K=2$, the average performance ratio of Algorithm 1 satisfies
\begin{align}
\lim_{n\to \infty} PR(G)&\geq \frac{p_1^2v_1+(p_2+p_1p_2)v_2}{(2p_2\hspace{-1pt}+\hspace{-1pt}2p_1^2\hspace{-1pt}+\hspace{-1pt}3p_1p_2)(\frac{v_1}{2}\hspace{-1pt}+\hspace{-1pt}(v_2\hspace{-1pt}-\hspace{-1pt}v_1)\frac{1-p_1}{2-p_1})}\nonumber\\
&\geq \frac{8}{9}\approx 88.9\%, \nonumber
\end{align}
and it attains the minimum if $\frac{v_2}{v_1}\to 1^+$ and $p_1=p_2=\frac{1}{2}$.
\end{proposition}

The proof is given in Appendix \ref{app:guarantee2} of the Supplementary Material of this TMC submission. This proposition suggests that the greedy matching’s average performance is surprisingly good (around $90\%$ of the optimum), which is much greater than $50\%$ in the worst case. It may be counter-intuitive that the average performance ratio of Algorithm 1 is the smallest when all edges have almost the same weights (not exactly the same), but this is actually consistent with the worst-case instance in Fig.~\ref{fig:systemmodel}. There we greedily choose only the middle edge of weight $1+\epsilon$ instead of the two side edges of total weight 2. As $\epsilon\rightarrow 0$, the greedy matching's performance worsens as compared to the optimum, which is equivalent to $\frac{v_2}{v_1}\rightarrow 1^+$ in Proposition \ref{pro:guarantee2}. As $p_1$ and $p_2$ get close to each other, the case that adjacent edges have nearly similar weights happens more frequently.

Similarly, for $K\geq 3$, we can obtain the lower bound for $PR(G)$ as a function of $V=\{v_1,v_2,\dots,v_K\}$ and $P=\{p_1,p_2,\dots,p_K\}$ and show that the greedy matching's average performance is always close to the optimum. Moreover, based on \eqref{equ:optimalupper} and the general formula in \eqref{equ:anyk}, we prove that the ratio $PR(G)$ is minimized when the possible weight values are similar given a uniform weight distribution $p_1=\cdots=p_K=\frac{1}{K}$.

\begin{proposition}\label{pro:uniformguarantee}
For an arbitrary $K$, if $p_1=\cdots=p_K=\frac{1}{K}$, the average performance ratio of Algorithm 1 in large-scale linear networks satisfies
\begin{align}
\lim_{n\to \infty} PR(G)&\geq \frac{\sum_{k=1}^{K} v_{k}\frac{(K-1)^{K-k}}{(K+1)^{K-k+1}}}{\sum_{k=1}^{K} v_{k} \frac{K}{(2K+1-k)(2K-k)}}\nonumber\\
&\geq 1-(\frac{K-1}{K+1})^K \geq 1-e^{-2}\approx 86.5\%,\nonumber
\end{align}
where the second inequality becomes equality when all the possible weight values become similar (i.e., $\frac{v_K}{v_1}\rightarrow 1^+$).
\end{proposition}

The proof is given in Appendix \ref{app:uniformguarantee} of the Supplementary Material of this TMC submission. The result here is consistent with Proposition \ref{pro:guarantee2}, and the average performance ratio of Algorithm 1 is the smallest when all edges have almost the same weights. The ratio slightly reduces as $K$ increases.

\section{Average-Case Analysis for D2D Sharing in 2D Grid Networks}\label{sec:grid}

In wireless networks, 2D grids are widely used to model social mobility of users (e.g., \cite{sensorgrid,gvgrid}). In this section, we analyze the average performance ratio and the parallel complexity of Algorithm 1 to validate its performance on planar user connectivity graphs. Note that the average-case analysis of 2D grids is an important benchmark for the more general random graphs analyzed in the following sections.

\subsection{Average Performance Analysis of Optimal Matching}\label{sec:generaloptimal}

It is infeasible to obtain the exact value of the average total weight $\mathbb{E}_{W}[f^\star(G,W)]$ under the optimal matching due to the exponential number of the possible matchings. Instead, we propose a method to compute an upper bound for the denominator $\mathbb{E}_{W}[f^\star(G,W)]$ in \eqref{equ:average} using a methodology that holds for general graphs. This upper bound will be used to derive a lower bound for the average performance ratio $PR(G)$ in \eqref{equ:average} later.

In any graph $G=(U,E)$, each matched edge $e_{ij}\in E$ adds value $w_{ij}$ to the final matching. Equivalently, we can think of it as providing individual users $u_i$ and $u_j$ with equal benefit $w_{ij}/2$. For any user $u_i$, this individual benefit does not exceed half of the maximum weight of its neighboring edges. Using this idea and summing over all users, the total weight of the optimal matching is upper bounded by
\begin{align}
f^\star(G,W)\leq \frac{1}{2}\sum_{u_i\in U} \max_{u_j\in A(u_i)} w_{ij}. \label{equ:newoptimal}
\end{align}
By taking expectation over the weight distribution, we obtain the closed-form upper bound of the average total weight. 

\begin{proposition}\label{pro:generalupper}
For a general graph $G=(U,E)$ with the weight set $V\hspace{-2pt}=\hspace{-2pt}\{v_1,v_2,\hspace{-1pt}\dots\hspace{-1pt},v_K\hspace{-1pt}\}$ and the weight distribution $P\hspace{-2pt}=\hspace{-2pt}\{p_1,p_2,\hspace{-1pt}\dots\hspace{-1pt},p_K\hspace{-1pt}\}$, the average total weight of the optimal matching is upper bounded by
\begin{align}
&\mathbb{E}_W[f^\star(G,W)]\leq \nonumber\\
&\frac{1}{2} \sum_{u_i\in U}\sum_{k=1}^K v_k((\sum_{i=1}^k p_i)^{|A(u_i)|}-(\sum_{i=1}^{k-1} p_i)^{|A(u_i)|}), \label{equ:generaloptimal}
\end{align}
where $|A(u_i)|$ is the cardinality of $A(u_i)$.
\end{proposition}

The proof is given in Appendix \ref{app:generalupper} of the Supplementary Material of this TMC submission.

\subsection{Average Performance Analysis of Algorithm 1}

We start with the probabilistic analysis of the parallel complexity of Algorithm 1. The result follows a similar reasoning as in the case of linear networks, but the proof is more subtle. This is because in the case of 2D grids, the number of possible chains that start from any given node $u_i$ and have non-decreasing weights is no longer two (toward left or right) as in $1\times n$ grid networks, but exponential in the size of the chain (since from each node there are $4-1=3$ `out' ways for the chain to continue), and such chains now form with non-negligible probability. This problem is not an issue for Algorithm 1 since every node will need to use priorities over ties among neighbors whose edge has the same weight. This significantly reduces the number of possible chains that are relevant to a user's decisions and we can prove the following proposition.
\begin{proposition} \label{pro:gridcomplexity}
In large-scale $n\times n$ grids, Algorithm 1 runs in $O(\log n)$ time w.h.p..
\end{proposition}

The proof is given in Appendix \ref{app:gridcomplexity} of the Supplementary Material of this TMC submission. In conclusion, our distributed matching algorithm has low complexity and provides a great implementation advantage compared to the optimal but computational-expensive centralized matching. 

\begin{figure}[t]\centering
\vspace{-0.2cm}
\includegraphics[width=\hsize]{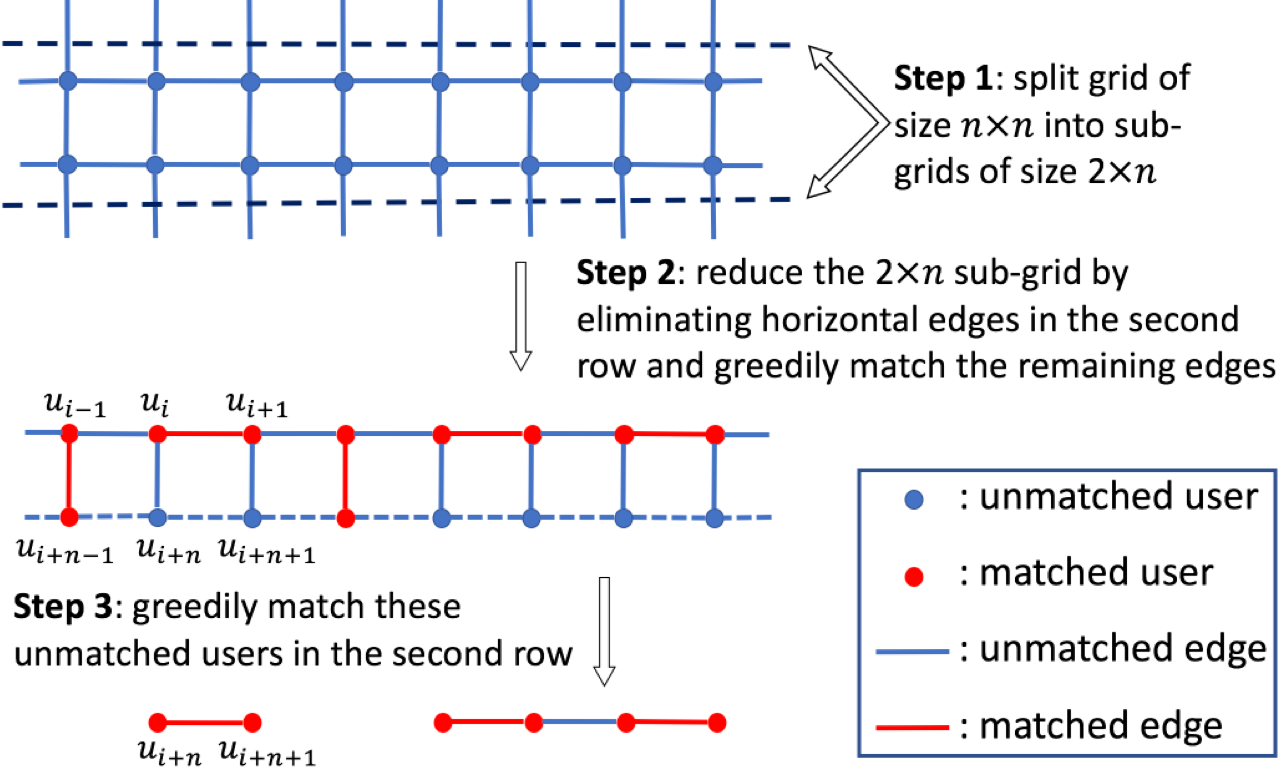}
\captionsetup{font=footnotesize}
\caption{Illustration of the grid reduction process in three steps for analyzing greedy matching. Algorithm 1 adds the red-colored edges to the greedy matching in each step.}
\label{fig:grid}
\vspace{-0.3cm}
\end{figure}

We next analyze the average total weight of the greedy matching, i.e., the numerator $\mathbb{E}_{W}[\hat{f}(G,W)]$ in \eqref{equ:average}. Unlike Section \ref{sec:greedy}, in the case of 2D grid networks we cannot directly use dynamic programming since matching users does not divide the grid into sub-grids. One may want to extend our previous result in linear networks to $n\times n$ grid, by dividing it into $n$ linear networks of size $1\times n$. However, this provides a poor lower bound because all the vertical edges become unavailable to match. Alternatively, we split the grid network into sub-grids in a way that keeps half of the vertical edges, and then estimate a tighter lower bound by further creating sub-graphs without cycles. Our procedure involves the following three steps (see Fig.~\ref{fig:grid}).

\vspace{0.1cm}
\noindent\emph{Step 1}: Split the $n\times n$ grid into $n/2$ sub-grids of size $2\times n$ by eliminating the corresponding vertical edges between sub-grids.

\vspace{0.1cm}
\noindent\emph{Step 2}: For each $2\times n$ sub-grid after step 1, eliminate all the horizontal edges in the second row (i.e., the blue dashed lines of the `Step 2' sub-graph in Fig.~\ref{fig:grid}) to create a graph without cycles. Then we analyze the greedy matching's performance over the remaining edges by using dynamic programming techniques.

\vspace{0.1cm}
\noindent\emph{Step 3}: For all the unmatched users in the second row, greedily match them by using the results in linear networks (see Section \ref{sec:greedy}).

To analyze the average total weight of the greedy matching, we first note that the graph created by step 2 can always be divided into two sub-graphs with the similar graph structure by matching an arbitrary edge. Here, a sub-graph with the similar graph structure refers to a sub-grid of smaller size $2\times n'$ (with any $0\leq n'<n$) and also with all the horizontal edges eliminated in the second row. Further, we can show that an edge that will certainly match in Algorithm 1 can be found within the first $2K$ edges (including the first $K$ horizontal edges in the first row and the corresponding $K$ vertical edges) of the created graph, by using the similar arguments as in Lemma~\ref{lem:firstk}\footnote{Here, without loss of generality, we assume that each user facing the same weights of adjacent edges assigns higher priority to match with the neighbor of smaller index. For example, the three neighbors $u_{i-1}$, $u_{i+1}$ and $u_{i+n}$ of user $u_i$ have decreasing priority to match when they have the same weight with $u_i$.}.

\begin{figure}[t]\centering
\vspace{-0.2cm}
\includegraphics[width=\hsize]{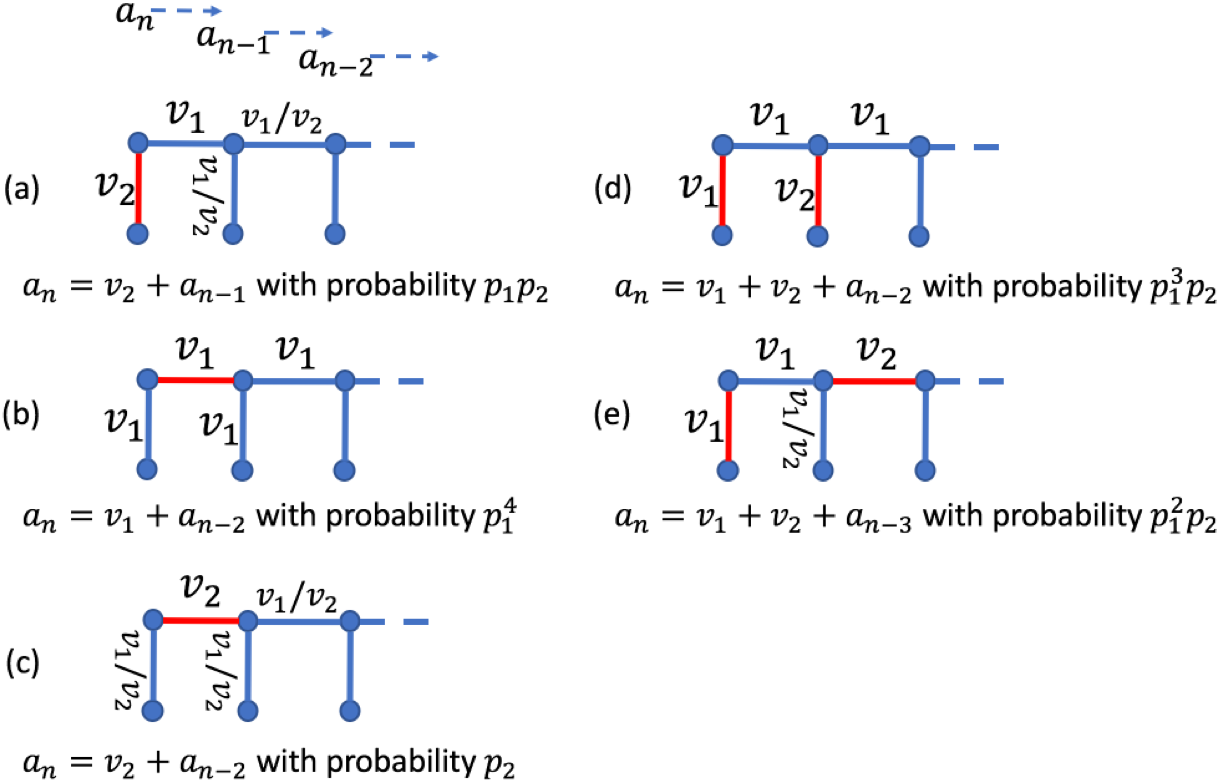}
\captionsetup{font=footnotesize}
\caption{Given the weight set size $K=2$ and $v_1<v_2$, an edge that certainly matches is always found within the first $2K=4$ edges, as marked in red. In each case, after matching the red edge, the remaining graph still has the similar structure but with a smaller size $n-1$, $n-2$ or $n-3$.}
\label{fig:gridexample}
\vspace{-0.3cm}
\end{figure}

Fig.~\ref{fig:gridexample} shows an illustrative example for $K=2$, and we always find such an edge (marked in red) with local maximum weight within the first $2K=4$ edges. This edge will be matched in Algorithm 1, and the remaining graph has the similar structure but with a smaller size. Then, by considering all the $K^{2K}$ weight combinations of the first $2K$ edges, we can similarly derive the recursive formula for the greedy matching as in linear networks. Note that in the example of $K=2$ in Fig.~\ref{fig:gridexample}, we reduce the totally $K^{2K}=16$ combination cases into $5$ cases ((a)-(e)) by combining these with the same certainly matched edges, and obtain the corresponding recursive formula for each of them. The final expected recursive formula is given by
\begin{align}
&a_n\hspace{-3pt}=\hspace{-3pt}(1\hspace{-2pt}-\hspace{-2pt}p_1^4)v_2\hspace{-2pt}+\hspace{-2pt}p_1^2v_1\hspace{-2pt}+\hspace{-2pt}p_1p_2a_{n-1}\hspace{-2pt}+\hspace{-2pt}(p_1^3+p_2)a_{n-2}\hspace{-2pt}+\hspace{-2pt}p_1^2p_2a_{n-\hspace{-1pt}3}.\nonumber
\end{align}
Based on this, we can similarly derive the general formula for $\{a_n\}$ when $n$ is large by using asymptotic analysis. Moreover, this method can also be extended for any possible weight distribution. 

Then, after the matching in Step 2, users in the second row form linear segments with different lengths in step 3 (see Fig~\ref{fig:grid}), and the greedy matching in these segments can be similarly analyzed as in Section \ref{sec:greedy}. Finally, we combine the analysis in steps 2 and 3 for the greedy matching's performance, and compare to the upper bound for the optimal matching in \eqref{equ:generaloptimal} to obtain the lower bound for $PR(G)$.

\begin{proposition}\label{pro:gridbound}
In large-scale $n\times n$ grids with the weight set $V=\{v_1=1,v_2=1+\Delta\}$ and uniform weight distribution $p_1=p_2=\frac{1}{2}$, the average performance ratio of Algorithm 1 satisfies 
\begin{align}
\lim_{n\to \infty} PR(G)\geq \frac{0.9213+0.6967\Delta}{1+0.9375\Delta},\nonumber
\end{align}
This ratio decreases from $92.1\%$ to $74.3\%$ when weight difference $\Delta$ increases from $0^+$ to $\infty$.
\end{proposition}

The proof is given in Appendix \ref{app:gridbound} of the Supplementary Material of this TMC submission. Different from Propositions \ref{pro:guarantee2} and \ref{pro:uniformguarantee} in linear networks, in the case of grids similar weight values (i.e., $\Delta\to 0^+$) no longer lead to the minimum average performance ratio. Intuitively, even if the instance in Fig.~\ref{fig:systemmodel} (three horizontal edges with similar weights) happens in grids, each user has at least a vertical neighbor to match. 

\begin{figure}[t]\centering
\vspace{-0.3cm}
\includegraphics[width=7cm]{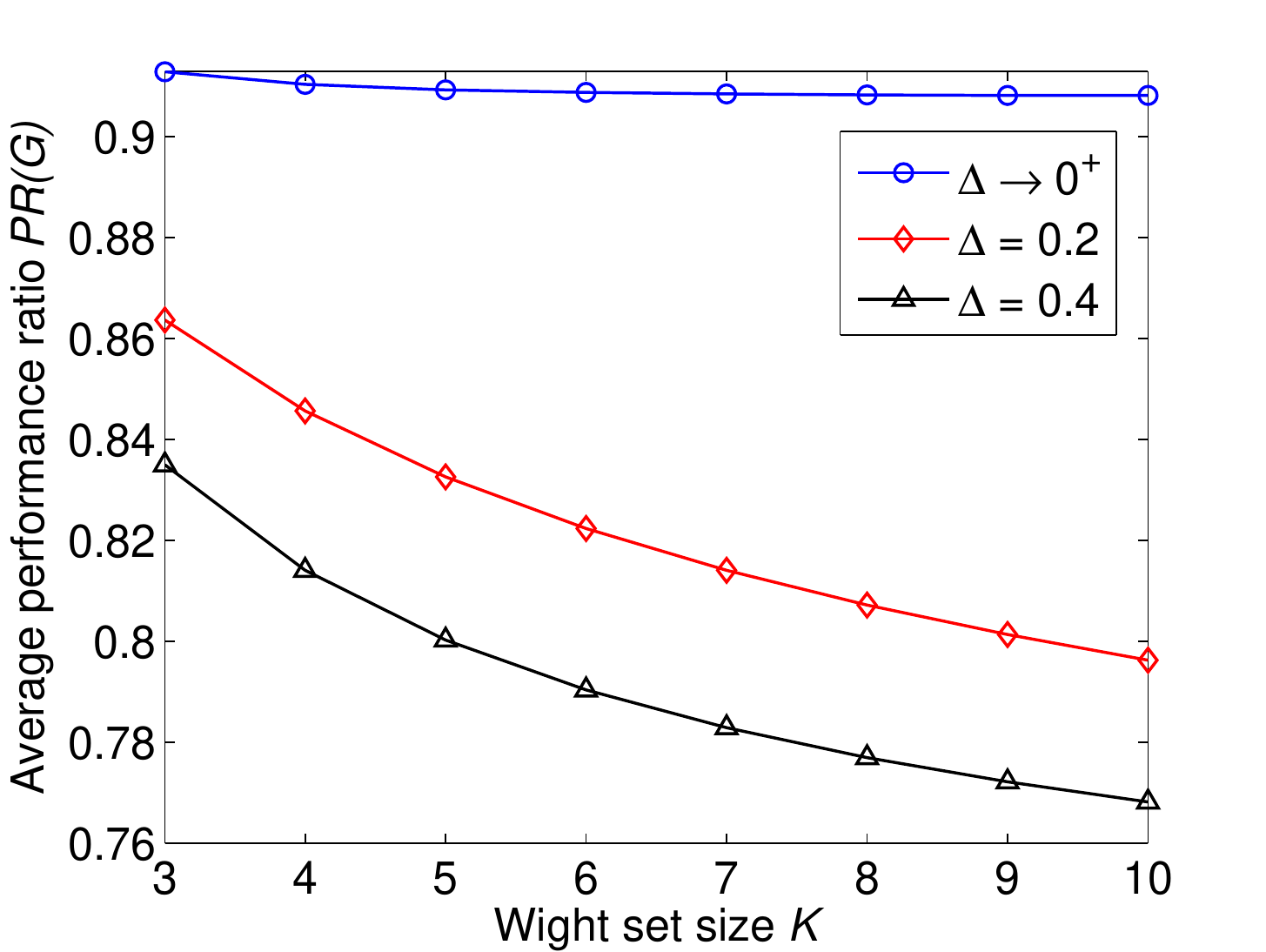}
\captionsetup{font=footnotesize}
\caption{The average performance ratio of Algorithm 1 in $n\times n$ grids versus the weight set size $K$ and weight difference $\Delta$. Here we assume edge weights are uniformly chosen from the weight set $V=\{1,1+\Delta,\dots,1+(K-1)\Delta\}$. }
\label{fig:arbk}
\vspace{-0.5cm}
\end{figure}

Finally, we further extend our analysis for larger weight set size $K>2$. In Fig.~\ref{fig:arbk}, we present the average performance ratio of Algorithm 1 in $n\times n$ grids against arbitrary $K$. Consistent with Proposition \ref{pro:gridbound}, here the average performance ratio bound decreases with $\Delta$ and is larger than $76\%$. It also decreases with $K$, which is also observed for linear networks.

\section{Average-Case Analysis for D2D Sharing in $G(n,p)$ Networks} \label{sec:random}

In practice, a mobile user may encounter a random number of neighbors. In this section, we extend our analysis to random networks $G(n, p)$, where $n$ users connect with each other with probability $p$ and hence each user has in the average (an order of magnitude) $d=np$ neighbors. Although the actual spatial distribution of users is not necessarily planar, such random graphs can still represent their connectivity on the ground and the analysis also holds.

We study the average performance ratio of Algorithm 1 in the cases of dense random graphs with a constant $p$ (i.e., dense since $d=np$ increases linearly in $n$) \cite{gnpbook2}, and sparse random graphs with a constant average neighbor number $d<1$ (i.e., $p<1/n$) \cite{gnpproperty}. Unlike the 2D grid networks, the structure of the random network $G(n, p)$ is no longer fixed due to the random connectivity. Though it is more technically difficult to analyze the average performance of Algorithm 1 for random graph structure, we are able to derive the ratio using statistical analysis in the two important cases below. For intermediate values of $d$ where our techniques cannot be applied, we have used exhaustive sets of simulations.

\subsection{Average-Case Analysis of Dense Random Graphs}

Given $p$ remains a constant, as $n$ increases, 
each user will have an increasing number of neighbors with the largest possible weight value $v_K$. Since such edges are preferred by greedy matching, as $n$ goes to infinity, the greedy matching will almost surely provide the highest possible total matching value of $nv_K/2$ ($n/2$ pairs of users with weight $v_K$).

\begin{proposition}\label{pro:performancegnp}
For a large-scale random graph $G(n,p)$ with a constant $p$, the average performance ratio of Algorithm 1 satisfies $PR=100\%$ w.h.p..
\end{proposition}

The proof is given in Appendix \ref{app:performancegnp} of the Supplementary Material of this TMC submission. In this result, we have taken expectation over both $G$ and $W$ in the definition of the average performance ratio $PR$. Note that the computation complexity is not anymore $O(\log n)$ in this case due to the increasing graph density. An obvious bound is $O(|E|)=O(n^2)$ proved in \cite{paper8}.

\subsection{Average-Case Analysis of Sparse Random Graphs}

In this subsection, we consider that the connection probability is $p=d/n$ and hence each user has a constant average number of neighbors $d(n-1)/n\to d$ as $n$ becomes large. We first prove low parallel complexity for Algorithm 1 as long as each user has a small enough number of neighbors to pair with that depends on the edge weight distribution.

\begin{proposition}\label{pro:complexitygnp}
For large-scale $G(n,d/n)$ type of networks, Algorithm 1 runs in $O(\log n)$ time w.h.p. if $d<2/\max\{p_1,p_2,\dots,p_K\}$.
\end{proposition} 

The proof is given in Appendix \ref{app:complexitygnp} of the Supplementary Material of this TMC submission. Note that this condition is always satisfied when $d<1$ because the weight probability $p_k\leq 1$ for any $k$. 

Next, we focus on studying the average performance ratio $PR$ for sparse random graphs $G(n,d/n)$. 
The average total weight of the optimal matching can be upper bounded by \eqref{equ:generaloptimal}, which works for any graph. Then, we only need to study the average total weight $\mathbb{E}_{G\sim G(n,d/n),W}[\hat{f}(G,W)]$ of the greedy matching. Note that when matching any graph $G$, we can equivalently view that the weight of any matched edge is equally split and allocated to its two end-nodes. Then we can rewrite the above expression as follows:
\begin{align}\label{equ:gnpsum}
\mathbb{E}_{G\hspace{-1pt}\sim G(n,\hspace{-1pt}d/n),W} [\hat{f}(G\hspace{-1pt},\hspace{-1pt}W)]\hspace{-1pt}=\hspace{-1pt}n\mathbb{E}_{G\hspace{-1pt}\sim G(n,d/n),\hspace{-1pt}W} [x_i(G\hspace{-1pt},\hspace{-1pt}W)],
\end{align}
where $x_i(G,W)$ is half of the weight of the matched edge corresponding to each user $u_i$ under the greedy matching.

We cannot use dynamic programming directly to compute the average weight $\mathbb{E}_{G\sim G(n,d/n),W} [x_i(G,W)]$ per user in \eqref{equ:gnpsum} since $G(n,d/n)$ may have loops and it cannot be divided into independent sub-graphs. Given that $n$ is large and assuming $d<1$, then graph $G(n,d/n)$, with very high probability, is composed of a large number of random trees without forming loops. In this case the matching weight $x_i(G,W)$ of user $u_i$ only depends on the connectivity with other users in the same tree. To analyze $x_i(G,W)$, we want to mathematically characterize such trees which turn out to be `small' because $d<1$. Note that, in $G(n,d/n)$, each user has $n-1$ independent potential neighbors, and its random neighbor number follows a binomial distribution $B((n-1),d/n)$ with mean $(n-1)d/n\rightarrow d$, as $n$ becomes large. This binomial distribution can be well approximated by the Poisson distribution $Poi(d)$ (with mean $d$). We define $T(d)$ as a random tree where each node in the tree gives birth to children randomly according to the Poisson distribution $Poi(d)$.
\begin{proposition}\label{pro:gnptree}
Given a sparse random network $G(n,d/n)$ with $d<1$ and sufficiently large $n$, 
the average matching weight of any node $u_i$ is well approximated by the average matching weight of the root node of a random tree $T(d)$, i.e.,
\begin{align}
\hspace{-11pt}\lim_{n\to \infty}\hspace{-3pt} \mathbb{E}_{G\hspace{-1pt}\sim G(n,\frac{d}{n}),W} [x_i(G\hspace{-1pt},\hspace{-2pt}W)]\hspace{-3pt}=\hspace{-2pt} \mathbb{E}_{T\hspace{-1pt}\sim T(d),W}[x_{root}(T\hspace{-1pt},\hspace{-2pt}W)\hspace{-1pt}].\label{equ:gnptree} 
\end{align}
\end{proposition}

The proof is given in Appendix \ref{app:gnptree} of the Supplementary Material of this TMC submission. We will show numerically later that the approximation in \eqref{equ:gnptree} yields trivial performance gap and remains accurate as long as $d\leq 10$. By substituting \eqref{equ:gnptree} into \eqref{equ:gnpsum}, we obtain approximately the average total weight $\mathbb{E}_{G\sim G(n,d/n),W}[\hat{f}(G,W)]$. Hence, it remains to derive the form of $\mathbb{E}_{T\sim T(d),W} [x_{root}(T,W)]$. Given the \emph{recursive nature} of trees, we are able to use \emph{dynamic programming}.

The root node may receive multiple proposals from its children corresponding to different possible edge weights in the set $\{v_1,v_2,\dots, v_K\}$, 
and will match to the one (of them) with the maximum weight. We define $y_k$, $k\in\{1,2,\dots,K\}$, to denote the probability that the root node receives a proposal from a child who connects to it with an edge of weight $v_k$. Then, by considering all the possible weight combinations of the root's children, we can compute the probability to match a child with any given weight, using the proposal probabilities $y_k$. In a random tree $T(d)$, given the root node is matched with one of its children, the remaining graph can be divided into several sub-trees which are generated from the grand-child or child nodes of the root node. In any case, a sub-tree starting with any given node has the similar graph structure and statistical property as the original tree $T(d)$. 
Thus, we are able to analytically derive the recursive equations for finding the proposal probabilities $\{y_k\}$ for the root node. 

\begin{proposition}\label{pro:gnptreeyk}
In the random tree $T(d)$, for any $k\in \{1,2,\dots,K\}$, the proposal probability $y_k$ from a child of edge weight $v_k$ to the root node is the unique solution to the following equation:
\begin{align}
\hspace{-5pt}y_k\hspace{-1pt}=\hspace{-1pt}e^{-\hspace{-1pt}(p_K\hspace{-1pt}+\hspace{-1pt}\sum_{j=k+1}^K y_jp_j) d}\sum_{i=0}^\infty \frac{(p_kd)^i(1\hspace{-2pt}-\hspace{-2pt}(1\hspace{-2pt}-\hspace{-2pt}y_k)^{i+1})}{(i+1)!y_k}.\label{equ:yk}
\end{align}
\end{proposition}

The proof is given in Appendix \ref{app:gnptreeyk} of the Supplementary Material of this TMC submission. Though not in closed form, we can easily solve \eqref{equ:yk} using bisection, and then compute the probability that the root node matches to a child with any given weight. Based on that we derive the average matching weight $\mathbb{E}_{T\sim T(d),W} [x_{root}(T,W)]$ of the root for \eqref{equ:gnptree} and thus $\mathbb{E}_{G\sim G(n,d/n),W}[\hat{f}(G,W)]$ in \eqref{equ:gnpsum}. Finally, by comparing with \eqref{equ:generaloptimal} under the optimal matching, we can obtain the average performance ratio of Algorithm 1.

\subsection{Numerical Results for Random Graphs}

Next, we conduct numerical analysis for sparse random graphs with $d<1$ and random graphs with finite $d\geq 1$. 
To do that by using analytic formulas, we need to approximate the random graph by random trees, and one may wonder if the approximation error is significant (when $d>1$). 
To answer this question, we consider large network size of $n = 10,000$, with edge weights uniformly chosen from the weight set $V=\{1,2\}$ (`low' and `high'). 
Our extensive numerical results show that the difference between the simulated average matching weight $\mathbb{E}_{G\sim G(n,\frac{d}{n}),W} [x_i(G,W)]$ and the analytically derived average matching weight $\mathbb{E}_{T\sim T(d),W} [x_{root}(T,W)]$ in the approximated tree $T(d)$ is always less than $0.05\%$ when $d<1$ and is still less than $1\%$ even for large $1\leq d\leq 10$. This is consistent with Proposition \ref{pro:gnptree}.

\begin{figure}[t]\centering
\vspace{-0.3cm}
\includegraphics[width=7cm]{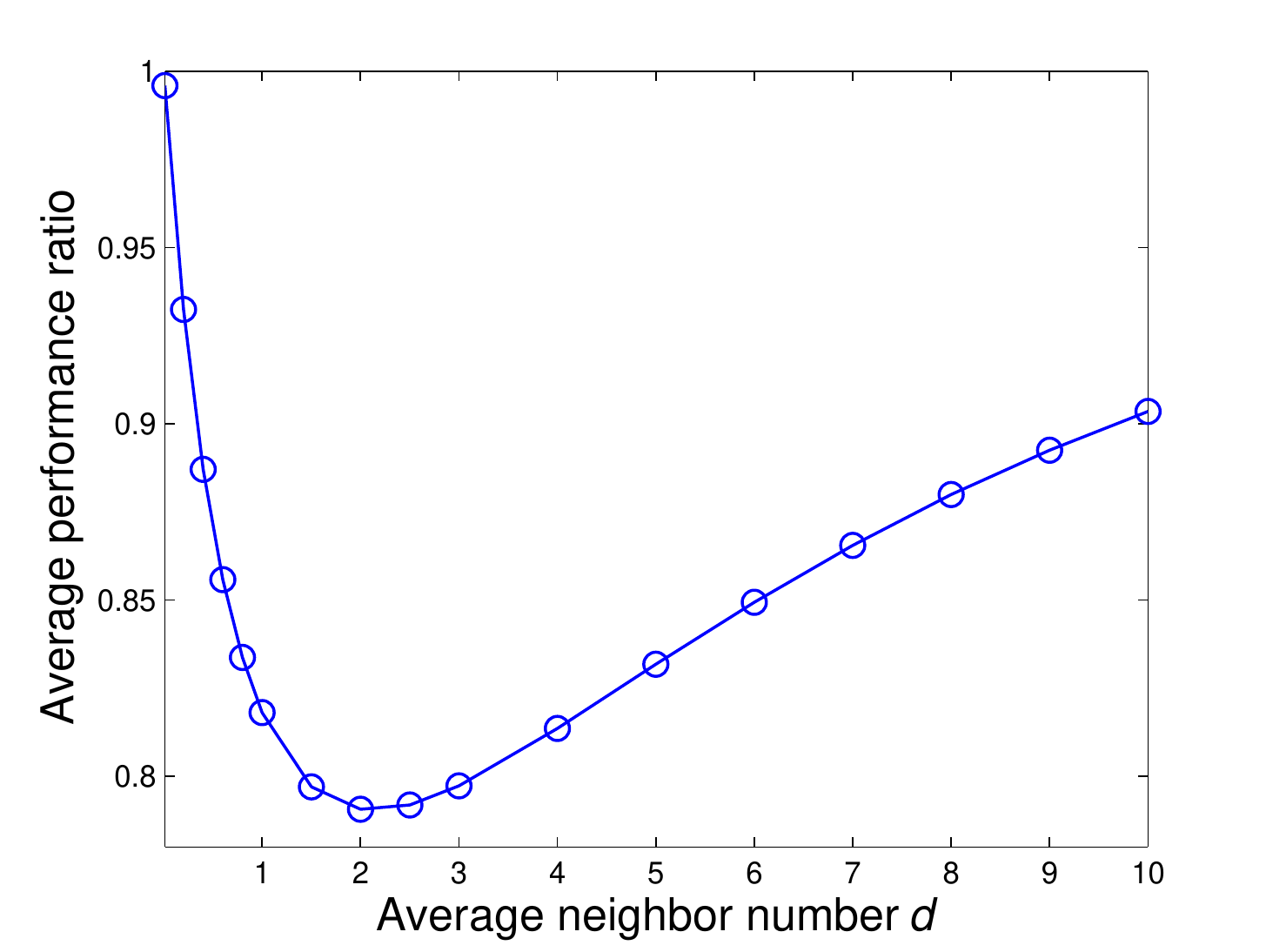}
\captionsetup{font=footnotesize}
\caption{The average performance ratio of Algorithm 1 in the large random graph $G(n,p=d/n)$ under different values of average neighbor number $d$.}
\label{fig:ratiognp}
\vspace{-0.5cm}
\end{figure}

Fig.~\ref{fig:ratiognp} shows the average performance ratio of Algorithm 1, which is greater than $79\%$ for any $d$ value. 
It approaches $100\%$ as $d$ is small in the sparse random graph regime. Intuitively, when the average neighbor number $d$ is small and users are sparsely connected, both Algorithm 1 and the optimal algorithm try to match as many existing pairs as possible, resulting in trivial performance gap. 
When $d$ is large, each user has many neighbors and choosing the second or third best matching in the greedy matching is also close to the optimum. This is consistent with Proposition \ref{pro:performancegnp} for dense random graphs. 

\section{Extension to Multi-unit D2D Resource Sharing} \label{sec:multiple}

In D2D sharing, a user may have multiple units of resources to supply or demand, and may share with multiple users at a time. For instance, an Android phone user may open up personal hotspot and share data connections with up to 10 users at the same time. In this section, we extend our average-case analysis to multi-unit resource sharing in linear networks. Though more involved, our analysis can also be extended to grid networks.

\subsection{Problem Description}
Similar to the single-unit linear network in Fig.~\ref{fig:line}, we consider a large-scale linear sharing network where each user $u_i$ locally connects with two adjacent users $u_{i-1}$ and $u_{i+1}$ and has $q_i$ units of resource demand (or supply) to share. Note that, in the multi-unit resource sharing, as long as we assume that each node cannot be both a consumer and a supplier at the same time (within one round), the direction of the edges is implied by the identity of the nodes. There is no need to change to directed graph modeling. We define $Q=\{q_i\}$ as the quantity vector for all $n$ users and suppose the quantity $q_i$ is IID for each user $u_i$. Similar to problem $\mathcal{P}_1$, we formulate the following multi-unit weighted allocation problem:

\begin{subequations}
\begin{align}
&\mathcal{P}_2:&\max \quad & \sum_{i=1}^{n-1} w_{i,i+1}x_{i,i+1}, &\nonumber\\
&&\text {s.t.} \quad & x_{i,i+1}\hspace{-2pt}+\hspace{-2pt}x_{i-1,i}\hspace{-2pt}\leq \hspace{-2pt}q_i, \forall i\hspace{-2pt}=\hspace{-2pt}2,3,\dots,n-1, \label{equ:multiple1}\\
&&& x_{1,2}\leq q_1,\quad x_{n-1,n}\leq q_{n}, \label{equ:multiple2}\\
&&& x_{i,i+1}\in \{0,1,2,\dots\},\quad\forall i=1,2,\dots,n-1, \nonumber
\end{align}
\end{subequations}
\noindent where constraints \eqref{equ:multiple1} and \eqref{equ:multiple2} ensure that the total amount of resources allocated to user $u_i$ is constrained by her desired quantity $q_i$.

Note that the direct extension of Algorithm 1 to solve the multi-unit problem above is to break each user $u_i$ with quantity $q_i$ into $q_i$ copies of one-unit users. However, this greatly increases the dimensionality of the problem (with the increased network size from $n$ to $\sum_{i=1}^n q_i$) and unnecessarily introduces competition between copies of the same user. To solve $\mathcal{P}_2$ efficiently, we make changes to the matching phase of Algorithm 1: every time an edge $e_i$ (between users $u_i$ and $u_{i+1}$) with the local maximum weight is found by the previous proposal phase, the allocation $x_i$ is no longer updated to 1, but increased by the minimum quantity $\min\{q_i,q_{i+1}\}$. Meanwhile, the quantities of $q_i$ and $q_{i+1}$ are decreased by the same amount. 

Note that each user still runs the steps of the revised algorithm based on local information, and will stop once her desired quantity is fully met or she sees no available neighbor. Thus, for each pair of users with the local maximum weight, at least one of them will fully satisfy her quantity in the matching phase and stop running the algorithm. The linear sharing network can be split due to any user's termination. Then, by using the similar arguments from single-unit case in Section \ref{sec:linear}, we prove the multi-unit version of Algorithm 1 still has sub-linear parallel complexity. 

\begin{lemma}
In large-scale linear networks of $n$ users, our revised Algorithm 1 for multi-unit resource sharing runs in $O(\log n)$ time w.h.p.. 
\end{lemma}

\subsection{Average Performance Analysis}

To study the average performance ratio of the multi-unit version of Algorithm 1 for solving problem $\mathcal{P}_2$, we also start with the average performance analysis of the optimal allocation. First note that, in any graph $G=(U,E)$, the optimal allocation for individual user $u_i$ (with the maximum weight $\sum_{j\in A(u_i)} x_{ij}w_{ij}$) is to allocate all her $q_i$ units of resource to the neighbors with the largest weights. As the allocation to any edge $e_{ij}$ is constrained not only by $q_i$ but also by $q_j$, the individual allocation weight of $u_i$ is upper bounded by
\begin{align}
\sum_{t=1}^{|A(u_i)|} w_{ij_t}x_{ij_t}. \label{equ:individual}
\end{align}
where $j_t$ is the neighbor of $u_i$ with the $t$-th largest weight (i.e., $w_{ij_1}\geq w_{ij_2}\geq\cdots\geq w_{ij_{|A(u_i)|}}$) and the allocation $x_{ij_t}$ assigned to pair $(u_i,u_{j_t})$ is computed as follows:
\begin{align} \label{equ:fijb}
    x_{ij_t}=\left\{
\begin{array}{ll}
\min\{q_i,q_{j_t}\}, \quad\quad\quad\quad\quad\quad\quad\quad\quad\,\text{ if } t=1, \\
\max\{0,\min\{q_i-\sum_{i=1}^{t-1}q_{j_i},q_{j_t}\}\},\text{ if } t\geq 2.
\end{array}
\right.   \end{align}

Remember that in single-unit case, we derive the upper bound in \eqref{equ:newoptimal} for the optimal matching based on the idea that each matched edge $e_{ij}$ can be viewed as providing individual users $u_i$ and $u_j$ with equal benefit $w_{ij}/2$. Using this idea and the maximum individual weight in \eqref{equ:individual}, we similarly derive the following upper bound of the total weight under the optimal allocation in any graph $G=\{U,E\}$.

\begin{lemma}
For a general graph $G\hspace{-1pt}=\hspace{-1pt}(\hspace{-1pt}U,\hspace{-1pt}E\hspace{-1pt})$, the total weight of the optimal allocation is upper bounded by
\begin{align}
f^\star(G,W,Q)\leq \frac{1}{2}\sum_{u_i\in U} \sum_{t=1}^{|A(u_i)|} w_{ij_t}x_{ij_t}. \label{equ:newoptimal2}
\end{align}
\end{lemma}

In particular, when $q_i=1$ for all user $u_i$, \eqref{equ:newoptimal2} degenerates to \eqref{equ:newoptimal}. Note that in \eqref{equ:newoptimal2} we need to take expectation over both weight $W$ and quantity $Q$ distributions to compute the average performance.

\begin{figure}[t]\centering
\vspace{-0cm}
\includegraphics[width=8.5cm]{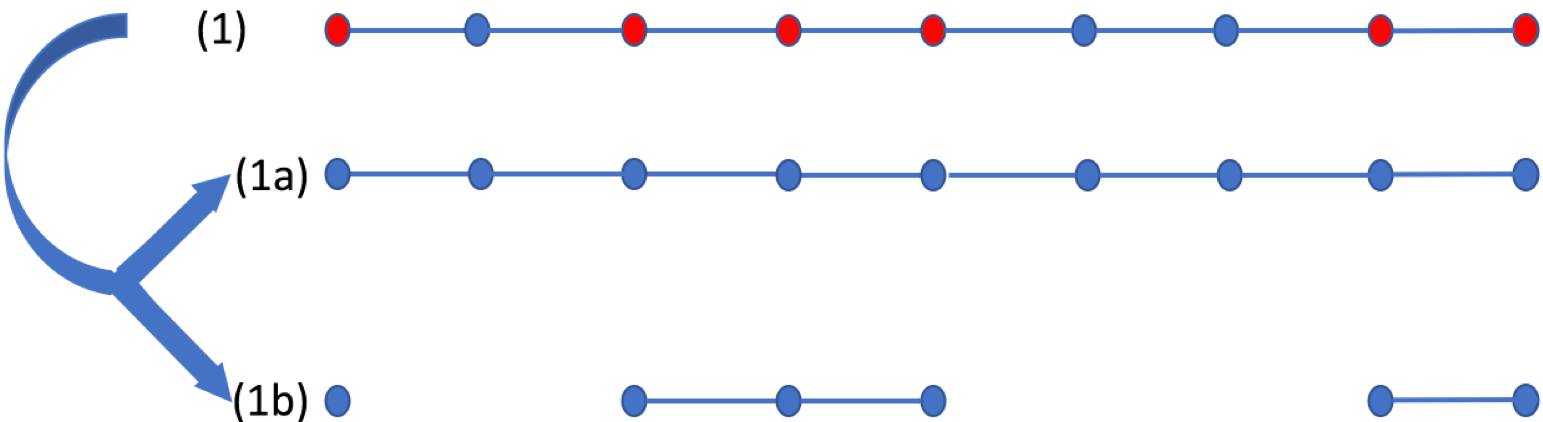}
\captionsetup{font=footnotesize}
\caption{Illustration of the new graph decomposition method for a multi-unit linear network example with two possible quantity values $1$ and $2$ for each user. All the red-colored nodes have quantity $2$ and all the blue-colored nodes have quantity $1$.}
\label{fig:multiple}
\vspace{-0.5cm}
\end{figure}

Next, to estimate a lower bound of the average total weight under the greedy allocation, we decompose the linear network into multiple single-unit linear networks. Different from the graph decomposition method proposed in Section \ref{sec:optimal}, here we decompose the graph based on user quantities instead of edge weights. As an illustration, Fig.~\ref{fig:multiple} shows the example for users with two possible quantity values ($1$ and $2$). We split it to a single-unit linear network in sub-graph (1a) in Fig.~\ref{fig:multiple} and the other sub-graph (1b) in Fig.~\ref{fig:multiple} to include the rest users with extra quantities. This new graph decomposition method helps us find a lower bound on the greedy allocation's performance.

Note that the average total weight of the single-unit greedy matchings in each sub-graph can be similarly analyzed as in Section \ref{sec:greedy}. Finally, by comparing the derived lower bound for the greedy allocation with the upper bound for the optimal allocation in \eqref{equ:newoptimal2}, we obtain the average performance ratio $PR(G)$. 

\begin{proposition}\label{pro:multiple}
In large-scale linear networks with edge weights and user quantities uniformly chosen from the set $V=\{1,1+\Delta\}$ and $Q=\{1,2\}$, the average performance ratio achieved by the multi-unit version of Algorithm 1 satisfies
\begin{align}
\lim_{n\to \infty}PR(G)\geq \frac{0.604+0.433\Delta}{0.75+0.5\Delta}.\nonumber 
\end{align}
This ratio increases with weight difference $\Delta$ and is larger than $80.5\%$ even when $\Delta\to 0^+$.
\end{proposition}

\begin{figure}[t]\centering
\vspace{-0.3cm}
\includegraphics[width=7cm]{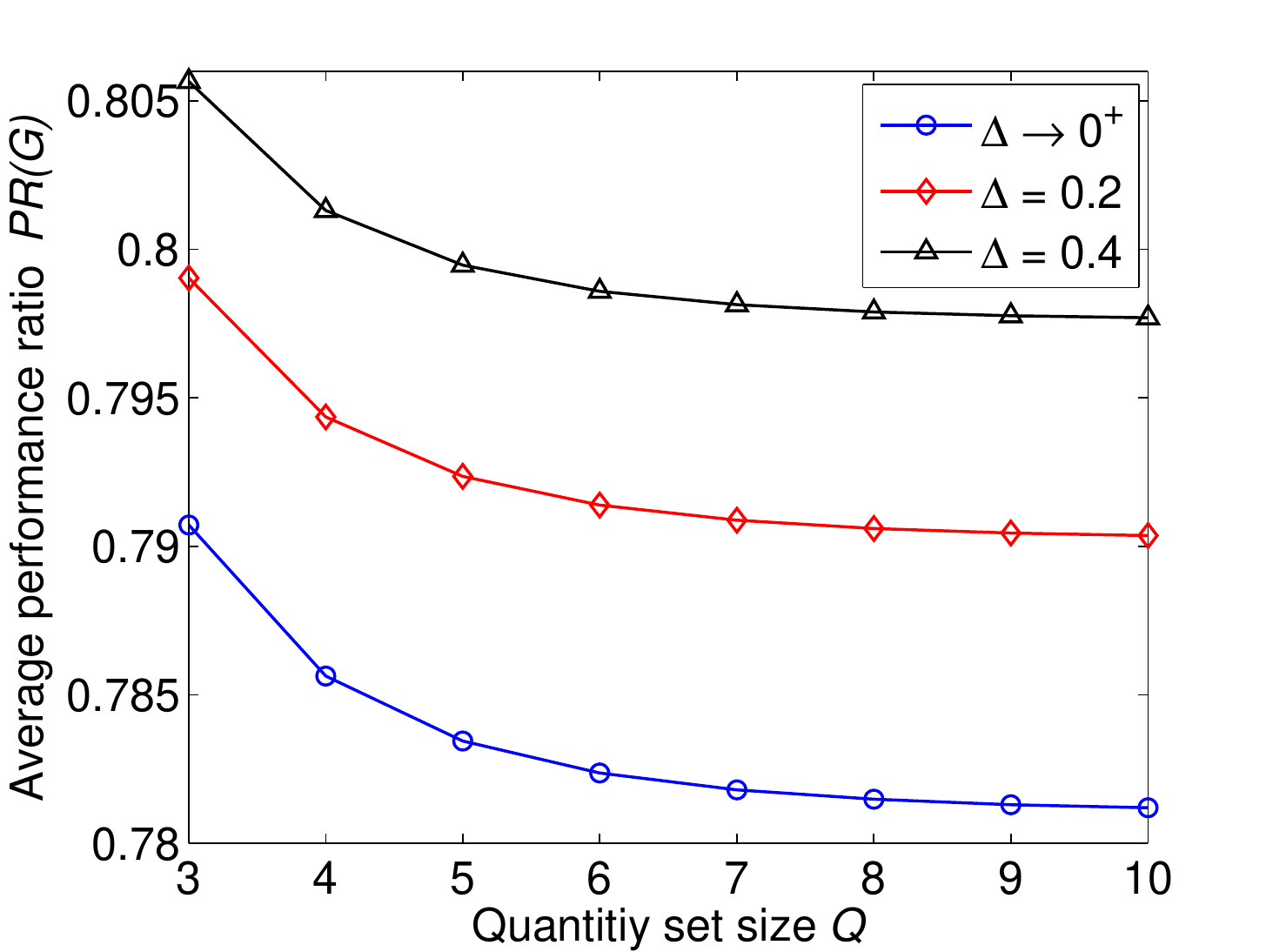}
\captionsetup{font=footnotesize}
\caption{The average performance ratio $PR(G)$ of the multi-unit version of Algorithm 1 in linear networks versus the quantity set size $Q$ under three different values of $\Delta$. Here we assume edge weights and user quantities are uniformly chosen from the sets $\{1,1+\Delta\}$ and $\{1,2,\dots,Q\}$, respectively.}
\label{fig:arbq}
\vspace{-0.5cm}
\end{figure}

The proof is given in Appendix \ref{app:multiple} of the Supplementary Material of this TMC submission. The obtained ratio increases with $\Delta$ and achieves its minimum value when all edges have the similar weights (i.e., $\Delta\to 0^+$). This is consistent with Proposition \ref{pro:guarantee2} for single-unit matching in linear networks. 

We also extend our analysis to any possible weight and quantity distributions. In Fig~\ref{fig:arbq}, we illustrate the average performance ratio $PR(G)$ of the multi-unit version of Algorithm 1 against different quantity set size $Q$. Similar to the case of $K=2$ in Proposition \ref{pro:multiple}, the lower bound for $PR(G)$ increases with $\Delta$ and is larger than $78\%$. It also decreases as $Q$ increases, as the performance gap is enlarged from the single-unit version.

\section{Practical Application Aspects}\label{sec:practical}
In practice, the network graphs that one may obtain by restricting the D2D sharing range may have different distributions than the 2D grids and the $G(n,p)$ graphs used in our analysis. In addition to that, the actual performance of the algorithm might be degraded because of communication failures of nodes that are far or mutual interference among pairs. In this section, we provide an investigation of the above issues. We construct a case study of collaborative caching in a network graph based on real data for mobile user locations. We check how well our analytical $G(n,p=d/n)$ performance measure in Section \ref{sec:caching} captures the actual performance of the greedy algorithm on the above realistic graph instances, by tuning $d$ to match the average number of neighbors in the instances. Later on, in Section \ref{sec:range}, we analyze the impact of D2D communication failures on the optimal selection of D2D maximum sharing range. Finally, in Section \ref{sec:interval}, we study the tradeoff in choosing $T$ (in minutes) for a dynamic scenario where users arrive/depart randomly and can participate in the sharing for several rounds.

\subsection{Case Study of D2D Caching}\label{sec:caching}

The $G(n,d/n)$ network studied in Section \ref{sec:random} assumes users connect with each other with the same probability $p=d/n$, and hence the average performance of Algorithm 1 in $G(n,d/n)$ is characterized by the average neighbor number $d$. However, in practice, the connectivity distribution of users can follow different laws due to the structure of the environment and the D2D communication limitations. To validate our analysis in scenarios of practical interest, we run our greedy matching algorithm on the D2D caching network corresponding to real mobile user data and compare the numerical results with our analytically derived results for $G(n,d/n)$ using Propositions \ref{pro:gnptree} and \ref{pro:gnptreeyk}.

We use the dataset in \cite{realdata} that records users' position information in a three-story university building. We choose three instances in the peak (in term of density) time from the dataset and each instance contains hundreds of users. For these users, we consider random local caching where users leverage short-range communications (e.g., Bluetooth) to share cached files following common interests \cite{caching}. We define the set of popular files as $F=\{1,2,\dots,10\}$ and each user $u_i$ caches three files from the library $F$ randomly. The individual caching file set for $u_i$ is denoted by $F_i\subset F$ with $|F_i|=3$. Any two users who cache different files (i.e., $F_i\neq F_j$), are allowed to share diverse files with each other as long as the distance between them is less than the range $L$ of the short-range communication. The corresponding weight (i.e., file sharing benefit) between them is determined by the number of different files they cache, i.e., $w_{ij}=|F_i\cup F_j|-|F_i\cap F_j|$.

\begin{figure}[t]\centering
\vspace{-0.3cm}
\includegraphics[width=7cm]{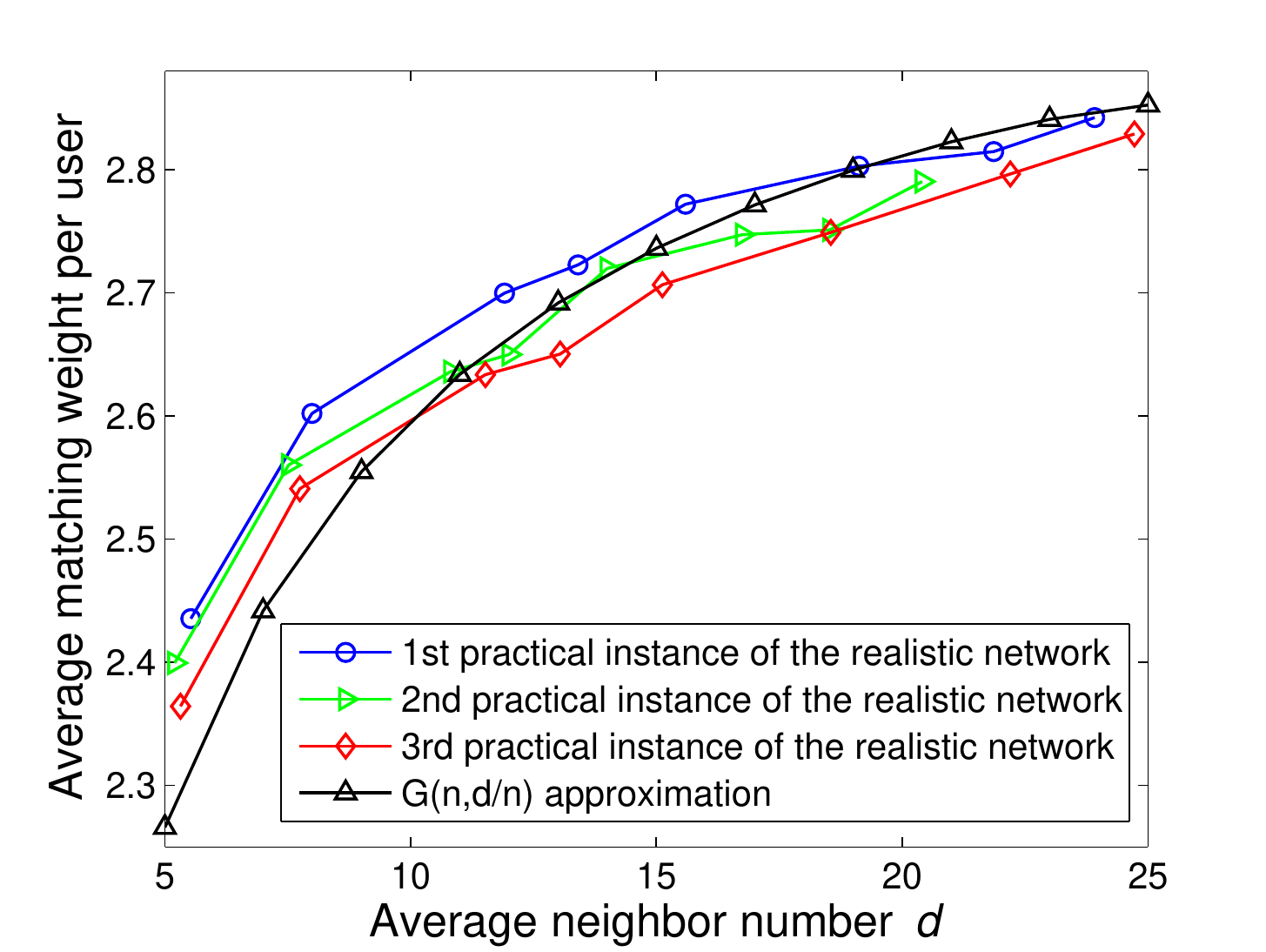}
\captionsetup{font=footnotesize}
\caption{The average matching weight per user of the greedy matching obtained by Algorithm 1 versus the average neighbor number $d$ in the three practical instances of the realistic network and the $G(n,p=d/n)$ network.}
\label{fig:realdataset2}
\vspace{-0.5cm}
\end{figure}

In the D2D caching network, by setting different values for $L$ the structure of the graph changes and the average number $d(L)$ of neighbors per user increases with $L$. In Fig.~\ref{fig:realdataset2}, we show the average matching weight (per user) of the greedy matching versus the average neighbor number $d$ for the three practical instances of the resulting user network and its $G(n,d/n)$ approximation. We observe that the average matching weight increases in $d$ since increasing $d$ (or increasing $L$) provides more sharing choices for each user. Our performance measure obtained for $G(n,d/n)$ approximates well the actual performance of our algorithm.

\subsection{D2D Sharing Range under Communication Failures}\label{sec:range}

Our numerical results from the previous section suggest that, as expected, the average matching weight of the greedy matching keeps increasing with the maximum D2D sharing range $L$. But this happens only because we did not include the deterioration of the quality of the D2D links when $L$ increases.
In fact, for two users who are connected and share resources via a D2D wireless link, a communication failure may occur more frequently due to the long-distance transmission or the mutual interference among different matched pairs. Such failures produce no matching and reduce the total matching weight. This suggests that after some value of $L$, the performance should decrease.

To show this tradeoff in choosing the best value for $L$, we assume there are two types of D2D communication failures: type-I failure caused by long-distance and type-II failure caused by interference. The transmission during resource sharing between any two users fails with a probability $\min\{1,\delta_1 D\}$ for type-I failure and a probability $\min\{1,\delta_2 I\}$) for type-II failure, where $D$ is the distance (in meters) between them and $I$ is the number of interfering pairs in proximity. $\delta_1$ and $\delta_2$ are scalar values representing the impact of distance and interference on the communication failures. The failure probability increases in the distance (based on a practical path-loss model) and the number of interfering pairs (see \cite{pathloss}). In our simulation experiment, we consider a large number $n=10,000$ of users uniformly distributed in a circular ground cell with a radius of $R=1000$ meters, and adjust the maximum sharing range $L$ in which two users implement D2D resource sharing.

\begin{figure}[t]\centering
\vspace{-0.3cm}
\includegraphics[width=7cm]{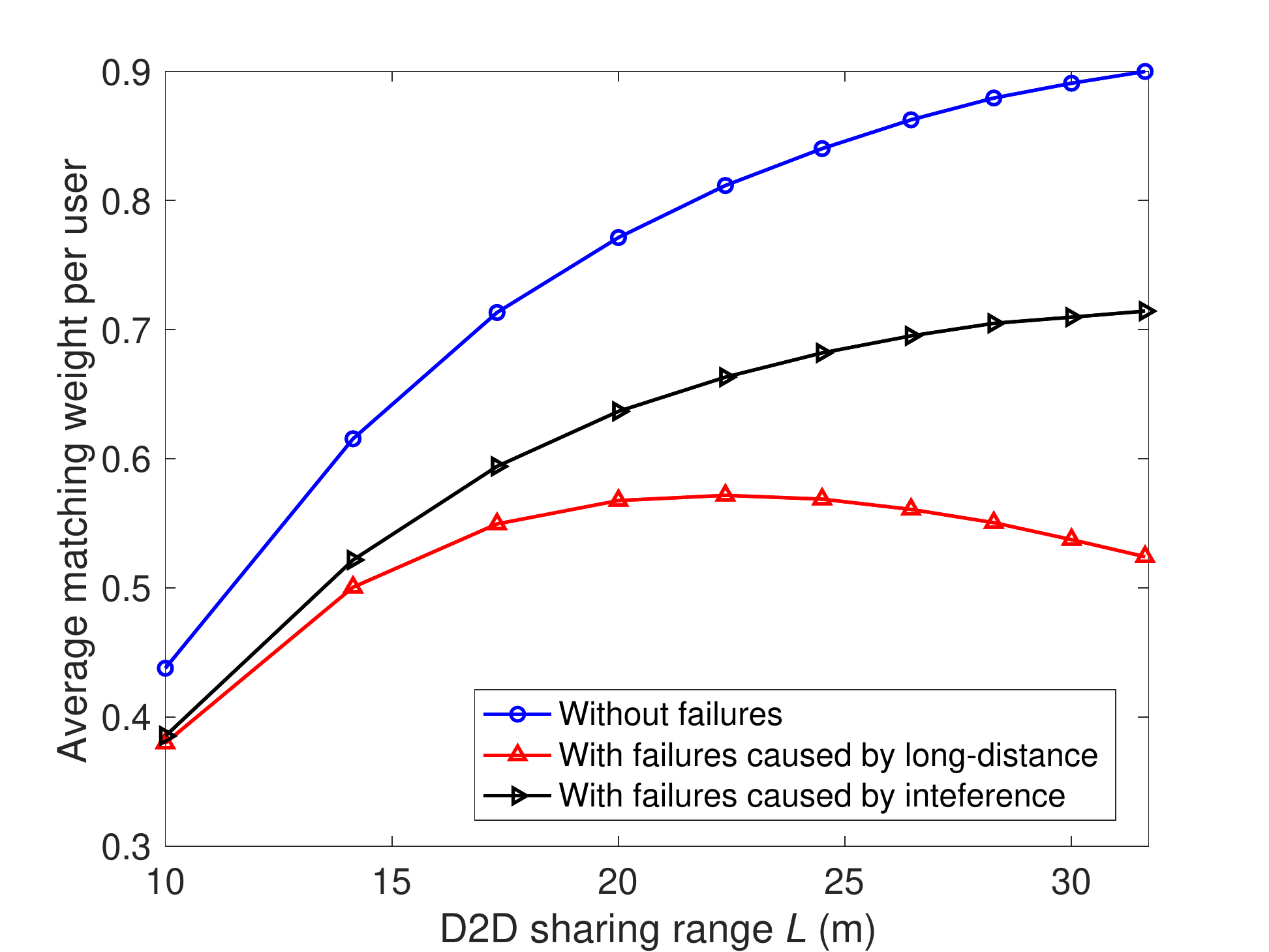}
\captionsetup{font=footnotesize}
\caption{The average matching weight per user of the greedy matching obtained by Algorithm 1 versus the maximum D2D sharing range $L$ for three cases: A) without D2D communication failures, B) with D2D communication failures caused by long-distance, and C) with D2D communication failures caused by interference. Here we set $\delta_1=0.02$ and $\delta_2=0.1$.}
\label{fig:greedy_real}
\vspace{-0.5cm}
\end{figure}

In Fig.~\ref{fig:greedy_real}, we depict the average matching weight (per user) of the greedy matching versus the maximum D2D sharing range $L$ for three cases: A) without D2D communication failures, B) with D2D communication failures caused by long-distance (type-I failure), and C) with D2D communication failures caused by interference (type-II failure). We observe that the performance gap between cases A and B increases with $L$, as well as in cases A and C. Intuitively, when $L$ is small, each user has few potential users to share resources or have interference, and failures occur rarely to be of an issue. But when $L$ is large, since most of the neighbors are located remotely and the channels between matched pairs may cross each other, there is a higher chance for the algorithm to choose a remote neighbor, in which case it incurs large path-loss (type-I failure) or interference (type-II failure). This is in contrast to the model without failures, where the performance of the system is always increasing in $L$.

\subsection{Optimal Time Interval} \label{sec:interval}

So far we have studied the average performance for a `one-shot' D2D resource sharing instance captured by the static graph $G$. In this subsection, we extend our model to a dynamic environment of running Algorithm 1 over multiple sharing rounds when users remain connected in the system for multiple rounds. An interesting question is how frequently to repeat Algorithm 1 to meet changes in the graph structure due to users' random arrivals, departures and movement. If the time interval $T$ (e.g., in minutes) to run the market matching algorithm is small, few new users will arrive and the existing users may not have extra resources to buy or sell. If $T$ is large, many arriving users may depart before any resource sharing happens because they don't want to wait for too long. Of course, the choice of $T$ depends on the type of resources that are being shared, how frequently a user can replenish its resources, if users remain active and connected to the application for a long time, etc. Hence our goal is not to estimate exactly the right $T$ since this is context-dependent, but to investigate the fundamental tradeoff between using small and large values for $T$.

\begin{figure}[t]\centering
\vspace{0cm}
\includegraphics[width=7cm]{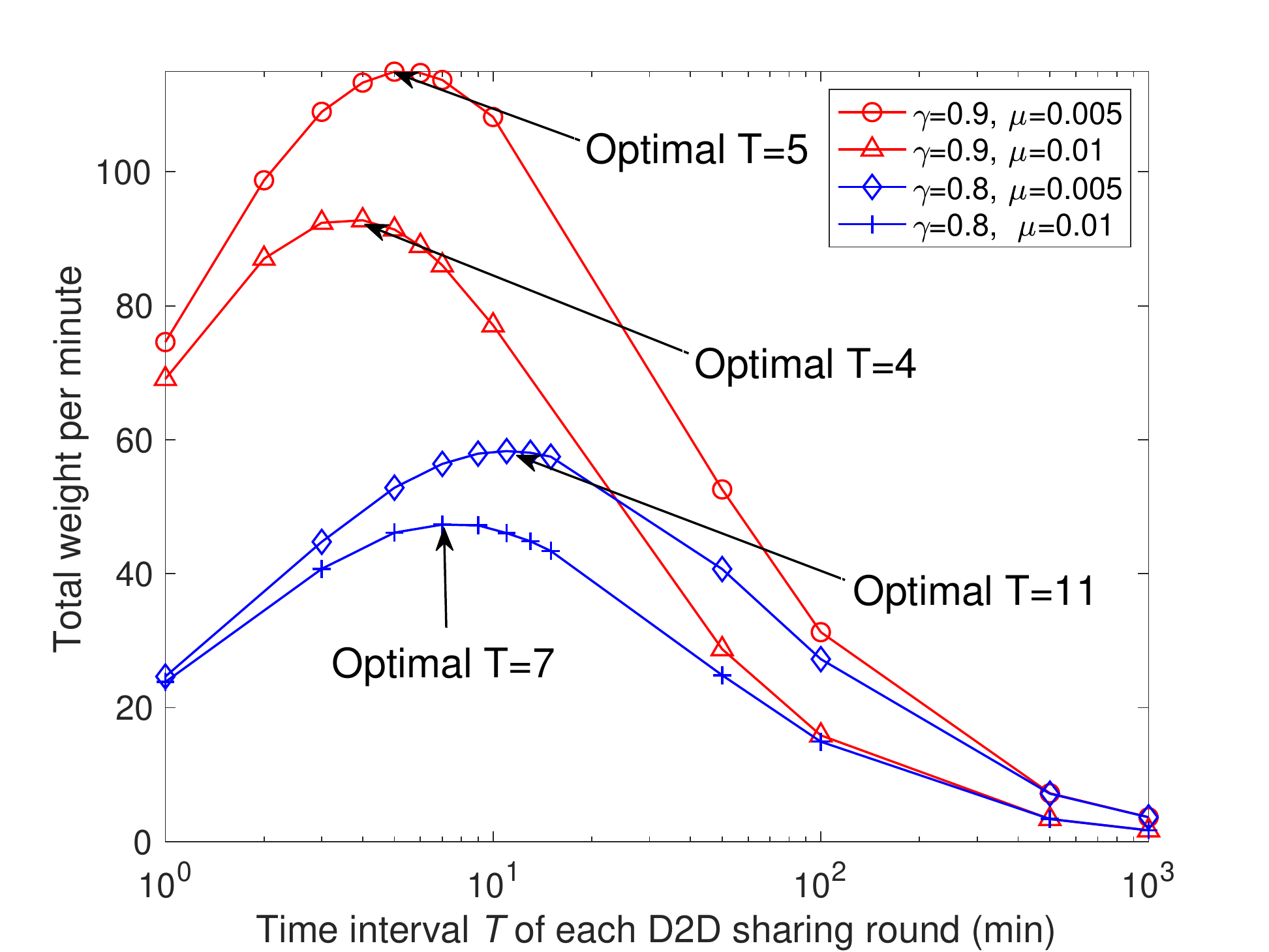}
\captionsetup{font=footnotesize}
\caption{Time-average total weight versus the time interval $T$ under different values of user sharing probability $\gamma$ and departure rate $\mu$. Here we assume users arrive with rate $\lambda=20$ per minute.}
\label{fig:frequency}
\vspace{-0.3cm}
\end{figure}

To characterize this tradeoff in system performance when choosing $T$, we consider a dynamic scenario that a fixed number $\lambda$ per minute of new users arrive for resource sharing in a circular ground cell with a radius of $R=1000$ meters, and each user will leave the network after an exponential random time with rate $\mu$. The maximum D2D sharing range $L$ is assumed to be $100$m between users. Furthermore, for those who stayed since the last round, they are still interested to share resources again in the new round with probability $\gamma<1$ and thus a user has resources to share for $1/(1-\gamma)$ rounds on average even if it remains in the system for a longer time. Let $M$ be the average number of active participants in the steady-state. In our experiment setting, we have $M=M\gamma e^{-\mu T}+\sum_{t=0}^{T-1}\lambda e^{-\mu(T-t)}$, where the first right-hand-side term of this equation tells the average number out of $M$ users from the last round to stay and share in the current round, and the second term tells the average number of new arriving users during the last period $T$.

We run Algorithm 1 in this dynamic scenario and Fig.~\ref{fig:frequency} shows the time-average total weight (per minute) versus $T$ under different values of $\gamma$ and $\mu$. It first increases and then decreases with $T$, since a small $T$ value does not allow enough new users to share, while a large $T$ value discourages many users who are impatient to wait. 
The optimal $T$ decreases with the probability $\gamma$ for each user to keep sharing, since more users are available for sharing due to a larger $\gamma$ and thus we can run the matching more frequently to minimize departures because of delayed service. The optimal $T$ decreases with departure rate $\mu$, as users are more likely to leave the network and we need a smaller $T$ to engage them in sharing.

\section{Conclusions}\label{sec:conclusions}
In this paper, we adopt a greedy matching algorithm to maximize the total sharing benefit in large D2D resource sharing networks. 
This algorithm is fully distributed and has sub-linear complexity $O(\log n)$ in the number of users $n$. 
Though the approximation ratio of this algorithm is $1/2$ (a worst-case result), we conduct average-case analysis to rigorously prove that this algorithm provides a significantly better average performance ratio compared to the optimum in large linear and grid networks. We then extend our analysis to random networks $G(n,p)$ and to multi-unit resource sharing. We also use real mobile user location data to show that our analytical $G(n,p)$ performance measure approximates well D2D networks encountered in practice. Finally, we consider the effect of communication failures due to increasing the communication range and study the related optimization problem. 

An interesting direction for future research is to consider the case of users staying for multiple rounds and being able to choose when to get matched. Users become strategic: accept a current matching or wait for a possibly better one in the future. The equilibrium strategies in such a system depend on parameters such as the distribution of matching values and the rate of arrivals and departures. Even if we restrict the topology of the graph to be linear, how do we analyze the resulting game? We also plan to extend our D2D resource sharing model that involves directly connected devices to multi-hop networks where intermediate devices can serve as `connectors' between the source and destination. One can capture the user connectivity graph with one-hop D2D connections and then add the edges when considering two-hop connections, etc.



%

\begin{IEEEbiography}[{\includegraphics[width=1in,height=1.25in,clip,keepaspectratio]{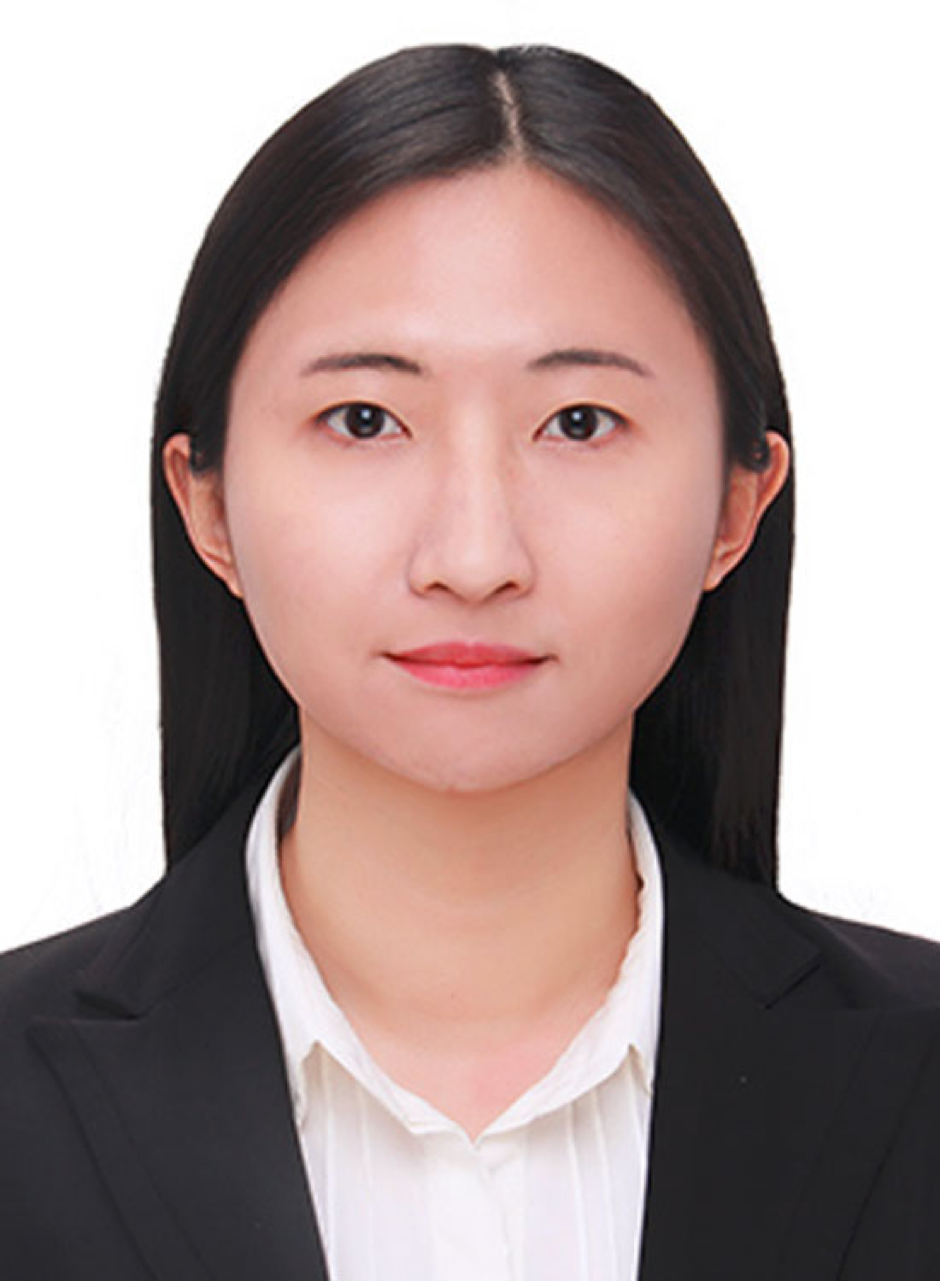}}]{Shuqin Gao} received the B.S. and M.S. degrees from Shanghai Jiao Tong University, China, in 2014 and 2017, respectively, and the Ph.D. degree from Singapore University of Technology and Design, Singapore, in 2022. Her research interests include network economics, mechanism design and performance analysis of distributed systems.
\end{IEEEbiography}

\vspace{-1cm}
\begin{IEEEbiography}[{\includegraphics[width=1in,height=1.25in,clip,keepaspectratio]{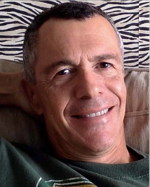}}]{Costas A. Courcoubetis} 
was born in Athens, Greece and received his Diploma (1977) from the National Technical University of Athens, Greece, in Electrical and Mechanical Engineering, his MS (1980) and PhD (1982) from the University of California, Berkeley, in Electrical Engineering and Computer Science. He was MTS at the Mathematics Research Centre, Bell Laboratories, Professor in the Computer Science Department at the University of Crete, Professor in the Department of Informatics at the Athens University of Economics and Business, Professor and Associate Head in the ESD Pillar, Singapore University of Technology and Design, and since 2021 Presidential Chair Professor in SDS, CUHK, Shenzhen. His current research interests include sharing economy and mobility, economics and performance analysis of networks and internet technologies, regulation policy, smart grids and energy systems, resource sharing and auctions. He received the 2022 MSOM Best OM Paper in Management Science Award and the 2021 MSOM Service Management SIG Best Paper Award.
\end{IEEEbiography}
\vspace{-1cm}
\begin{IEEEbiography}[{\includegraphics[width=1in,height=1.25in,clip,keepaspectratio]{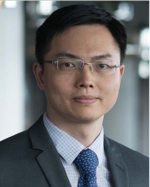}}]{Lingjie Duan} (S'09-M'12-SM'17) received the Ph.D. degree from The Chinese University of Hong Kong in 2012. He is an Associate Professor of Engineering Systems and Design with the Singapore University of Technology and Design (SUTD). In 2011, he was a Visiting Scholar at University of California at Berkeley, Berkeley, CA, USA. His research interests include network economics and game theory, cognitive and green networks, and energy harvesting wireless communications. He is an Editor of IEEE Transactions on Wireless Communications. He was an Editor of IEEE Communications Surveys and Tutorials. He also served as a Guest Editor of the IEEE Journal on Selected Areas in Communications Special Issue on Human-in-the-Loop Mobile Networks, as well as IEEE Wireless Communications Magazine. He received the SUTD Excellence in Research Award in 2016 and the 10th IEEE ComSoc Asia-Pacific Outstanding Young Researcher Award in 2015.
\end{IEEEbiography}

\clearpage

\appendices

\section{Proof of Proposition \ref{pro:complexitylinear}}\label{app:complexitylinear}

For any user $u_i$ in the linear network, let $H(u_i)$ denote the length of the longest chain (sequence of edges) that has non-decreasing weights and starts from $u_i$ towards the left or right side. Let $I(u_i)$ be the indicator variable that $H(u_i)$ is greater than $c\log n$ for some constant $c$. Let $I(G)$ be the indicator variable that the linear graph $G$ has at least one such chain with length greater than $c\log n$. Then we have
\begin{align}
\mathbb{E}(I(G))\leq\mathbb{E}(
\sum_{i=1}^n I(u_i))\leq 2nq, 
\end{align}
where $q$ denotes the probability that $c\log n$ consecutive edges has non-decreasing weights.

The weight of each edge is assumed to independently take value from $K$ kinds of weight values $\{v_1,v_2,\dots,v_K\}$ according to the probability distribution $P=\{p_1,p_2,\dots,p_K\}$. Then, for any $c\log n$ edges, there are totally:
\[\sum_{k=1}^{K-1}{{1+c\log n}\choose k}{K-2 \choose k-1},\]
kinds of non-decreasing weight combinations, and for each of the combinations, the probability to happen is upper bounded by $(\max\{p_1,p_2,\dots,p_K\})^{c\log n}$. Therefore, the upper bound of the probability $q$ that any $c\log n$ consecutive edges have non-decreasing weights is given by:
\[q\leq (\max\{p_1,p_2,\dots,p_K\})^{c\log n}\sum_{k=1}^{K-1} {1+c\log n \choose k}{K-2 \choose k-1}\]
\[\leq n^K (\max\{p_1,p_2,\dots,p_K\})^{c\log n}.\]
Then, we have:
\[\mathbb{E}(I(G)) \leq 2n^{K+1}(\max\{p_1,p_2,\dots,p_K\})^{c\log n}.\]
Note that when $c>\frac{K+1}{-\log \max\{p_1,p_2,\dots,p_K\}}$, $\mathbb{E}(I(G))$ converges to $0$ when $n\to \infty$.

Therefore, we can conclude that, in $1\times n$ grid, the algorithm run in $O(\log n)$ time w.h.p. according to the first moment method.

\section{Proof of Proposition \ref{pro:upperoptimal}}\label{app:upperoptimal}

In the first step of the graph separation method, we create a new graph by reducing the weights of all edges by the smallest possible weight value $v_1$ and the graph's weight vector is simplified to $W'=W-v_1=\{w_1-v_1,w_2-v_1,\dots,w_n-v_1\}$ with zero weight for some $w_i=v_1$ (if any). First note that the optimal matching of the graph with weight vector $W=\{w_1,w_2,\dots,w_n\}$ may no longer be the optimal matching of the graph with weight vector $W'=W-v_1=\{w_1-v_1,w_2-v_1,\dots,w_n-v_1\}$, and thus we have $\sum_{i=1}^{n}(w_i-v_1)x^\star_i(W)=\sum_{i=1}^{n}w_i'x^\star_i(W)\leq \sum_{i=1}^{n}w_i'x^\star_i(W') =f^\star(W')$ where $x^\star_i(W)$ is the optimal matching indicator for edge $e_i$ in a graph with weight vector $W$. Thus, the relationship between $f^\star(W)$ and $f^\star(W')$ satisfies the following inequality
\begin{align}
f^\star(W)&\hspace{-2pt}=\hspace{-2pt}\sum_{i=1}^{n}x^\star_i(W)w_i\hspace{-1pt}=\hspace{-1pt}v_1\sum_{i=1}^{n}x^\star_i(W)\hspace{-2pt}+\hspace{-2pt}\sum_{i=1}^{n}(w_i\hspace{-1pt}-\hspace{-1pt}v_1)x^\star_i(W)\nonumber\\
&\leq v_1 \frac{n}{2}+\sum_{i=1}^{n}(w_i-v_1)x^\star_i(W)\nonumber\\
&\leq v_1 \frac{n}{2}+f^\star(W'),
\end{align}
where the first inequality uses the fact that $\sum_{i=1}^{n}x^\star_i(W)\leq \frac{n}{2}$. This is because that, for any linear network with $n$ users, no matter what its weight vector $W=(w_1,w_2,\dots,w_n)$ is, the total number of edges in the optimal matching of the graph is $\lceil\frac{n}{2}\rceil$ at most and we ignore the ceiling operation since we consider a very large $n$. After taking expectation, the average total weights of the optimal matchings in $W$ and $W'$ satisfies
\begin{align}
\mathbb{E}_{W}[f^\star(W)]\leq v_1 \frac{n}{2}+\mathbb{E}_{W}[f^\star(W')].\label{equ:step1}
\end{align}

In the graph with weight vector $W'=W-v_1$, the weight $w_i'\hspace{-2pt}=\hspace{-2pt}w_i\hspace{-1pt}-\hspace{-1pt}v_1$ of an edge $e_i$ whose original weight $w_i\hspace{-2pt}=\hspace{-2pt}v_1$ now becomes $0$ and such edge appears with probability $\Pr(w_i=v_1)=p_1$. We remove all these edges as they have no influence on computing the optimal matching. Then, the graph becomes a lot of linear segments with different length as in Fig.~\ref{fig:optimaldeduct} and the average total number of remaining edges is $n(1-p_1)$.

Note that each segment starts from an edge with nonzero weight, and the following edge has zero weight with probability $p_1$ and nonzero weight with probability $1-p_1$. Thus, a segment has length $1$ with probability $p_1$ and has length larger than $1$ with probability $1-p_1$. Similarly, a segment of length $t$ needs $t-1$ following consecutive nonzero-weight edges and then ends with a zero-weight edge. Thus, the probability that a segment has length $t$ is $p_1(1-p_1)^{t-1}$. For any nonzero-weight edge, the probability that this edge is within a segment of length $t$ now is $\frac{tp_1(1-p_1)^{t-1}}{\sum_{k=1}^{+\infty} kp_1(1-p_1)^{k-1}}=tp_1^2(1-p_1)^{t-1}$. Accordingly, the average number of edges that are in a segment with length $t$ is given by the average total number of nonzero-weight edges in the graph with weight vector $W'$ times the probability, i.e., $n(1-p_1)\times tp_1^2(1-p_1)^{t-1}$. Further, the average number of segments with length $t$ is $np_1^2(1-p_1)^{t}$.

For a segment of length $t$, at most $\lceil\frac{t}{2}\rceil$ edges are included in any possible matching. Thus, for the graph with weight vector $W'$, the average total number of nonzero-weight edges included in the optimal matching $\mathbb{E}_{W}[\sum_{i=1}^n x_i^{\star}(W')]\leq \sum_{t=1}^{+\infty} np_1^2(1-p_1)^{t}\lceil\frac{t}{2}\rceil=n\frac{1-p_1}{2-p_1}$. Now, we further deduct $v_2-v_1$ from the weights of all nonzero-weight edges in the second step of the graph separation method to create another new graph with weight vector $W''=(W'-(v_2-v_1))^+$ and obtain that
\begin{align}
\mathbb{E}_{W}[f^\star(W')]\leq (v_2-v_1)\frac{1-p_1}{2-p_1}n+\mathbb{E}_{W}[f^\star(W'')].\label{equ:step2}
\end{align}
where the inequality is also because the total weight of the optimal matching in graph with $W''$ must be larger than or equal to the total weight of any possible matching.

If weight space size $K=2$, by aggregating the inequalities (\ref{equ:step1}), (\ref{equ:step2}) and the fact that $f^\star(W'')=0$ as $W''=(W-v_2)^+=0$, we finally obtain the upper bound of $\mathbb{E}_{W}[f^\star(W)]$ in the case of $K=2$ as follows: 
\[\mathbb{E}_{W}[f^\star(W)]\leq v_1\frac{n}{2}+(v_2-v_1)\frac{1-p_1}{2-p_1}n.\]

For a graph with weight set size $K$ larger than $2$, similar to the case of $K=2$ with $\{v_1,v_2\}$, we first deduct the weights of all edges with nonzero weight by $v_1$ in step 1 and by $v_2-v_1$ in step 2. For the graph created in step 2 with weight vector $W''=(W-v_2)^+$, the edges with zeros weight appear with probability $p_1+p_2$. Then, we can also remove all the edges with zero weight in this graph, and prove that the average total number of nonzero-weight edges included in the optimal matching, $\sum_{i=1}^n x_i^{\star}(W'')$ must be less than or equal to $n\frac{1-(p_1+p_2)}{2-(p_1+p_2)}$ using the similar arguments. Moreover, similarly, we further deduct $v_3-v_2$ from the weights of all nonzero-weight edges to create another new graph with weight vector $W'''=(W-v_3)^+$ and prove that
\begin{align}
&\mathbb{E}_{W} [f^\star(W')]\leq (v_3-v_2)n\frac{1-p_1-p_2}{2-p_1-p_2}+\mathbb{E}_{W} [f^\star(W''')].\nonumber
\end{align}
By repeating such deducting procedure for $K$ steps in the separation method, we can finally create a new graph with zero weight vector $(W-v_K)^+=0$ and the relationship between the total weight of the optimal matching in the original graph with weight vector $W$ and the total weight of the optimal matching in the decomposed graphs is
\begin{align}
&\mathbb{E}_{W} [f^\star(W)]\leq v_1\frac{n}{2}\hspace{-2pt}+\hspace{-2pt} (v_2\hspace{-2pt}-\hspace{-2pt}v_1)n\frac{1\hspace{-2pt}-\hspace{-2pt}p_1}{2\hspace{-2pt}-\hspace{-2pt}p_1} \hspace{-2pt}+\hspace{-2pt} (v_3\hspace{-2pt}-\hspace{-2pt}v_2)n\frac{1\hspace{-2pt}-\hspace{-2pt}p_1\hspace{-2pt}-\hspace{-2pt}p_2}{2\hspace{-2pt}-\hspace{-2pt}p_1\hspace{-2pt}-\hspace{-2pt}p_2}\hspace{-2pt}+\hspace{-2pt}\nonumber\\
&\cdots+ (v_{K}-v_{K-1})n\frac{1-\sum_{i=1}^{K-1}p_i}{2-\sum_{i=1}^{K-1}p_i} + \mathbb{E}_{W} [f^\star((W-v_K)^+])\nonumber\\
&=v_1 \frac{n}{2}+n\sum_{k=1}^{K-1} (v_{k+1}-v_k)\frac{1-\sum_{i=1}^k p_i}{2-\sum_{i=1}^k p_i}.\nonumber
\end{align}
The proof is completed.

\section{Proof of Proposition \ref{pro:generalanyk}}\label{app:generalanyk}
In the case of $K=2$, the recurrence formula of the average total weight under the greedy matching is given by:
\[a_n=p_1^2v_1+(p_2+p_1p_2)v_2+(p_2+p_1^2)a_{n-2}+p_1p_2a_{n-3},\]
and in the case of $K=3$, we also have
\begin{align}
&a_n=p_1(p_2p_3+p_1)v_1+p_2(p_1+p_2)(p_1+1)v_2\nonumber\\
&+p_3(p_1p_2+p_1+p_2+1)v_3+(p_1^2+p_2^2+p_1p_2+p_3)a_{n-2}\nonumber\\
&+(p_1p_2^2+p_1p_3+p_1^2p_2+p_2p_3)a_{n-3}+p_1p_2p_3a_{n-4}.\nonumber
\end{align}
Thus, we can imply that, for any arbitrary $K$, the recurrence formula of sequence $\{a_n\}$ has the following structure:
\begin{align}
a_n=&\beta_1v_1+\beta_2v_2+\cdots+\beta_K v_K\nonumber\\
&+\gamma_1a_{n-2}+\gamma_2a_{n-3}+\cdots+\gamma_{K}a_{n-K-1},\nonumber
\end{align}
where $\gamma_i$ denotes the probability that edge $e_i$ is the first edge must be added in Algorithm 1 (i.e., the first edge satisfying $w_i\geq w_{i+1}$ and $w_i>w_{i-1}$), and $\beta_k$ denotes the probability that an edge with weight $v_k$ is added because we add the first edge must be added (e.g., edge $e_i$ is the first edge must be added and edges $e_{i-2}$, $e_{i-4}$, and so on will also be added).

To obtain the general formula, we have to study the expression of $\gamma_i$ and $\beta_k$. First note that, given edge $e_i$ is the first edge must be added and $w_i=v_j$ (probability is $Pr(w_i=v_j)=\frac{1}{K}$), then we have $w_1<w_2<w_3\cdots<w_{i-1}<v_j$ (probability is $(\frac{1}{K})^{i-1} {{j-1}\choose{i-1}}$) and $w_{i+1}\geq v_j$ (probability is $j\frac{1}{K}$). Thus, we have
\[\gamma_i=\sum_{j=i}^K j(\frac{1}{K})^{i+1} {{j-1}\choose{i-1}}.\]

Next, we focus on finding the expression of $\beta_k$. We first consider the case of $k=K$ and $\beta_K$ denote the probability that an edge with the largest weight $v_K$ is added because we add the first edge must be added. Note that, if an edge $e_i$ is added because of adding the first edge must be added, then the first edge must be edge $e_{i}$ itself or an edge that is in position after $e_i$ and has higher weight than $e_{i}$. Since $v_K$ is the largest weight, an edge $e_i$ with weight $v_K$ will be added if and only if itself is the first edge must be added, i.e. $w_1<w_2<\cdots<w_{i-1}<w_i=v_K\geq w_{i+1}$ (the probability is $(\frac{1}{K})^{i}{{K-1}\choose{i-1}}$). Thus, $\beta_K$ is given by
\begin{align}
\beta_K&=\sum_{i=1}^{K}(\frac{1}{K})^{i}{{K-1}\choose{i-1}}=\frac{1}{K}(1+\frac{1}{K})^{K-1}.\nonumber
\end{align}
Similarly, we consider the case of $k=K-1$. Note that, if an edge $e_i$ with weight $v_{K-1}$ is not the first edge must be added but it is added because of adding the first edge, then the first edge must be in one of the position $i+2$, $i+4$ or above. In such a case, we must have $w_{i+2}>w_{i+1}>w_i=v_{K-1}$. But this is impossible since we only have one possible weight value larger than $v_{K-1}$. Thus, similarly, an edge $e_i$ with weight $v_{K-1}$ will be added if and only if itself is the first edge must be added, i.e., $w_1<w_2<\cdots<w_{i-1}<w_i=v_{K-1}\geq w_{i+1}$ (the probability is $(\frac{1}{K})^{i}{{K-2}\choose{i-1}}\frac{K-1}{K}$). Accordingly, $\beta_{K-1}$ is given by
\begin{align}
\beta_{K-1}&=\sum_{i=1}^{K-1}(\frac{1}{K})^{i}{{K-2}\choose{i-1}}\frac{K-1}{K}=\frac{K-1}{K^2}(1+\frac{1}{K})^{K-2}.\nonumber
\end{align}
Different from the above two cases, an edge $e_i$ with weight $v_{K-2}$ can be added not only when itself is the first edge must be added, i.e., $w_1<w_2<\cdots<w_{i-1}<w_i=v_{K-2}\geq w_{i+1}$ (the probability is $(\frac{1}{K})^{i}\frac{K-2}{K}{{K-3}\choose{i-1}}$), but also when the edge $e_{i+2}$ is the first edge must be added and $w_{i+1}=v_{K-1}$ and $w_{i+2}=v_{K}$, i.e., $w_1<w_2<\cdots<w_{i-1}<w_i=v_{K-2}<w_{i+1}=v_{K-1}<w_{i+2}=v_K\geq w_{i+3}$ (the probability is $(\frac{1}{K})^{i}{{K-3}\choose{i-1}}(\frac{1}{K})^2$). Thus, $\beta_{K-2}$ is given by
\begin{align}
\beta_{K\hspace{-2pt}-\hspace{-2pt}2}&\hspace{-2pt}=\hspace{-2pt}\sum_{i=1}^{K-2}(\frac{1}{K})^{i}{{K\hspace{-2pt}-\hspace{-2pt}3}\choose{i\hspace{-2pt}-\hspace{-2pt}1}}(\frac{K\hspace{-2pt}-\hspace{-2pt}2}{K}\hspace{-2pt}+\hspace{-2pt}\frac{1}{K^2})\hspace{-2pt}=\hspace{-2pt}\frac{(K\hspace{-2pt}-\hspace{-2pt}1)^2}{K^3}(1\hspace{-2pt}+\hspace{-2pt}\frac{1}{K})^{K\hspace{-2pt}-\hspace{-2pt}3}.\nonumber
\end{align}
In general, we can obtain the expression of $\beta_{K-t}$ for any $t\geq K$ as follows: 
\begin{align}
\beta_{K-t}&=\sum_{i=1}^{K-t}(\frac{1}{K})^{i}{{K-t-1}\choose{i-1}}(\frac{1}{K}(K-t)\nonumber\\
&+(\frac{1}{K})^3(\sum_{j=1}^{t-1}(j+K-t+1){{j}\choose 1})\nonumber\\
&+(\frac{1}{K})^5(\sum_{j=3}^{t-1}(j+K-t+1){{j}\choose 3})+\cdots)\nonumber\\
&=(1+\frac{1}{K})^{K-t-1}\frac{1}{K} (\frac{1}{K}(K-t)\nonumber\\
&+\sum_{i=3,5,\dots,K+1}(\frac{1}{K})^{i}(\sum_{j=i-2}^{t-1}(j+K-t+1){{j}\choose{i-2}})),\label{equ:ck}
\end{align}
where the second term can be simplified as follows:
\begin{align}
&\sum_{i=3,5,\dots,K+1}(\frac{1}{K})^{j}(\sum_{j=i-2}^{t-1}(j+K-t+1){{j}\choose{i-2}})\nonumber\\
&\overset{\text{(c)}}=\sum_{j=1}^{t-1}\sum_{i=3,5,\dots,j+2}(\frac{1}{K})^{i}(j+K-t+1){{j}\choose{i-2}}\nonumber\\
&=\sum_{j=1}^{t-1}\sum_{i=1,3,\dots,j}(\frac{1}{K})^{i+2}(j+K-t+1){{j}\choose{i}}\nonumber\\
&\overset{\text{(d)}}=\sum_{j=1}^{t-1}(\frac{1}{K})^{2}(j+K-t+1) \sum_{i=1}^{+\infty}\frac{(-1)^i+1^i}{2}(\frac{1}{K})^{i+2}{{j}\choose{i}}\nonumber\\
&=(1-\frac{1}{K})^t-\frac{K-t}{K},\nonumber
\end{align}
where (c) exchanges the integration order and (d) is based on the formula $\sum_{i=1}^n ia^i{n \choose i}=na(1+a)^{n-1}$. By substituting the above equation back into (\ref{equ:ck}), we obtain
\begin{align}
\beta_{K-t}&=(1+\frac{1}{K})^{K-t-1}\frac{1}{K}(1-\frac{1}{K})^t\nonumber\\
&=(1+\frac{1}{K})^{K}\frac{(K-1)^t}{(K+1)^{t+1}},\label{equ:alpha}
\end{align}
and it can be rewritten as
\begin{align}
\beta_k=\frac{(K-1)^{K-k}}{K^K(K+1)^{-k+1}}.\nonumber
\end{align}
The proof is completed.

\section{Proof of Proposition \ref{pro:guarantee2}} \label{app:guarantee2}
In the case that $k=2$, we have the following recurrence formula for $\{a_n\}$:
\begin{align}
a_n=p_1^2v_1+(p_2+p_1p_2)v_2+(p_2+p_1^2)a_{n-2}+p_1p_2a_{n-3}.\label{equ:greedyrecurrencnew}
\end{align}
Note that a general solution to (\ref{equ:greedyrecurrencnew}) is given by
\begin{align}
a_n&=\frac{p_1^2v_1+(p_2+p_1p_2)v_2}{2(p_2+p_1^2)+3p_1p_2}n+b,\nonumber\\
&=\frac{p_1^2v_1+(p_2+p_1p_2)v_2}{2+p_1p_2}n+b,\label{equ:general2} 
\end{align}
where $b$ can be any constant value. However, in our problem with $a_1=0$, $a_2=p_1v_1+p_2v_2$, the first two terms of the sequence do not follow an arithmetic progression. To obtain the converged $b$ as $n$ goes to infinity, we first list the following equations based on (\ref{equ:greedyrecurrencnew}):
\begin{align}
&a_1=0;\nonumber\\
&a_2=p_1v_1+p_2v_2;\nonumber\\
&a_3=p_1^2v_1+(p_2+p_1p_2)v_2+(p_2+p_1^2)a_1;\nonumber\\
&a_4=p_1^2v_1+(p_2+p_1p_2)v_2+(p_2+p_1^2)a_2+p_1p_2a_1;\nonumber\\
&\hdots\nonumber\\
&a_{n-1}=p_1^2v_1+(p_2+p_1p_2)v_2+(p_2+p_1^2)a_{n-3}+p_1p_2a_{n-4};\nonumber\\
&a_n=p_1^2v_1+(p_2+p_1p_2)v_2+(p_2+p_1^2)a_{n-2}+p_1p_2a_{n-3}.\nonumber
\end{align}
By summing them up, we obtain that
\begin{align}
&a_n+a_{n-1}+p_1p_2a_{n-2}=\nonumber\\
&\quad\quad\quad\quad \quad (p_1^2v_1+(p_2+p_1p_2)v_2)(n-2)+p_1v_1+p_2v_2.\label{equ:sum}
\end{align}
Therefore, by substituting (\ref{equ:general2}) into (\ref{equ:sum}), we can show that $b$ converges to $\frac{(2+p_1p_2)(p_1v_1+p_2v_2)-3(p_1^2v_1+(p_1+p_1p_2)v_2)}{(2+p_1p_2)^2}$ as $n\to \infty$. We finally obtain the general formula for the sequence $\{a_n\}$ in the limit of large enough network size as follows:
\begin{align}
&\lim_{n\to\infty} \mathbb{E}_{W}[\hat{f}(W=\{w_1,w_2\dots,w_n\})] \nonumber\\
&= \lim_{n\to\infty} a_n=\frac{p_1^2v_1+(p_2+p_1p_2)v_2}{2p_2+2p_1^2+3p_1p_2}n+O(1),\label{equ:generalformula2}
\end{align}

Based on (\ref{equ:optimalupper}) and (\ref{equ:generalformula2}), the average performance guarantee of Algorithm 1 is given by
\begin{align}
\lim_{n\to\infty}PR(n)\geq \frac{p_1^2v_1+(p_2+p_1p_2)v_2}{(2p_2+2p_1^2+3p_1p_2)(\frac{v_1}{2}+(v_2-v_1)\frac{1-p_1}{2-p_1})},\label{equ:boundk2}
\end{align}

Then, by substituting $p_2=1-p_1$ into (\ref{equ:boundk2}), we have
\begin{align}
\lim_{n\to\infty}PR(G)\geq \frac{p_1^2v_1+(1-p_1^2)v_2}{(2+p_1-p_1^2)(\frac{p_1}{4-2p_1}v_1+\frac{1-p_1}{2-p_1}v_2)},\label{equ:boundk2new}
\end{align}
which is a function of $v_1$, $v_2$ and $p_1$. Note that when $v_2$ and $p_1$ are fixed, this function is decreasing in $v_1$ since the ratio of the coefficients of $v_1$ in the numerator and the denominator $\frac{p_1^2}{\frac{p_1}{4-2p_1}}$ is less than or equal to the ratio of the constant terms in the numerator and the denominator $\frac{1-p_1^2}{\frac{1-p_1}{2-p_1}}$ for any $0\leq p_1\leq 1$. Moreover, the intrinsic bound for $v_1$ is $0\leq v_1<v_2$. Therefore, the average performance guarantee must achieve the minimum value when $v_1$ infinitely approaches $v_2$, i.e., $v_1\to v_2^-$. Now, given $v_1\to v_2^-$, (\ref{equ:boundk2new}) can be further rewritten as
\begin{align}
\lim_{n\to\infty}PR(G)\geq \frac{2}{(2+p_1-p_1^2)}=\frac{2}{\frac{9}{4}-(p_1-\frac{1}{2})^2}.\nonumber
\end{align}
This guarantee achieves its minimum value $\frac{8}{9}$ when $p_1=\frac{1}{2}$ obviously.

Thus, we can conclude that the average performance guarantee given in (\ref{equ:boundk2new}) achieves its minimum value $\frac{8}{9}$ when $\frac{v_2}{v_1}\to 1^+$ and $p_1=p_2=\frac{1}{2}$.
The proof is completed.

\section{Proof of Proposition \ref{pro:uniformguarantee}}\label{app:uniformguarantee}
For any general weight set size $K$, we can similarly derive the general formula of sequence $\{a_n\}$ as follows
\begin{align}
\lim_{n\to\infty}a_n= n\frac{\beta_1v_1+\beta_2v_2+\cdots+\beta_K v_K}{2\gamma_1+3\gamma_2+\cdots+(K+1)\gamma_{K}}+O(1).\label{equ:general}
\end{align}
By substituting $\beta_k=\frac{(K-1)^{K-k}}{K^K(K+1)^{-k+1}}$ and $\gamma_k=\frac{1}{K^{k+1}} \sum\limits_{i=k}^K i {{i-1}\choose{k-1}}$ into \eqref{equ:general}, we obtain the following formula:
\[\lim_{n\to\infty}a_n=n\sum_{k=1}^{K} v_{k}\frac{(K-1)^{K-k}}{(K+1)^{K-k+1}}+O(1).\]
Moreover, in Proposition \ref{pro:upperoptimal}, we prove that the average total weight under the optimal matching is upper bounded as follows
\begin{align}
\mathbb{E}_{W}[f^\star(W)]&\geq v_1\frac{n}{2}+n\sum_{k=1}^{K-1} (v_{k+1}-v_k)\frac{1-\sum_{i=1}^k p_i}{2-\sum_{i=1}^k p_i} \nonumber\\
&=n\sum_{k=0}^{K-1} v_{k} \frac{K}{(2K+1-k)(2K-k)},\nonumber
\end{align}
Thus, the average performance guarantee is given by:
\begin{align}
PR(G)\geq \frac{\sum_{k=1}^{K} v_{k}\frac{(K-1)^{K-k}}{(K+1)^{K-k+1}}}{\sum_{k=1}^{K} v_{k} \frac{K}{(2K+1-k)(2K-k)}}.\label{equ:bound}
\end{align}
Note that the ratio of the coefficients of $v_k$ in the numerator and the denominator increases as $k$ increases, i.e.,
\[\frac{\frac{(K-1)^{K-k}}{(K+1)^{K-k+1}}}{\frac{K}{(2K+1-k)(2K-k)}}\hspace{-1pt}<\hspace{-1pt}\frac{\frac{(K-1)^{K-k-1}}{(K+1)^{K-k}}}{\frac{K}{(2K-k)(2K-k-1)}},\forall k\in\hspace{-1pt}\{1,2,\hspace{-1pt}\dots\hspace{-1pt},K-1\}.\]
Thus, the guarantee given in (\ref{equ:bound}) is minimized when $v_1$ infinitely approaches $v_2$, $v_2$ infinitely approaches $v_1$ and so on. Moreover, the minimum value is given by
\[PR(G)\geq \frac{\sum_{t=0}^{K-1} \frac{(K-1)^t}{(K+1)^{t+1}}}{\sum_{t=0}^{K-1} \frac{K}{(K+1+t)(K+t)}}= 1-(\frac{K-1}{K+1})^K,\]
which is decreasing in $K$ when $K\geq 2$ and the limit is given by:
\[\lim_{K\to\infty}\hspace{-1pt}1-(\frac{K-1}{K+1})^K\hspace{-2pt}=\hspace{-2pt} \lim_{K\to\infty} 1-(1-\frac{2}{K+1})^{\frac{K+1}{2}\frac{2K}{K+1}}\hspace{-2pt}=\hspace{-2pt} 1-e^{-2}.\]
Moreover, note that the coefficients of $v_K$ in the numerator and the denominator are the same. Thus, the guarantee given in (\ref{equ:bound}) is maximized when $0\leq v_1<v_2<\cdots<v_{K-1}\ll v_K$ and the maximum value is given by
\[PR(G)\geq \frac{\frac{1}{K+1}}{\frac{1}{K+1}}= 1.\]
The proof is completed.

\section{Proof of Proposition \ref{pro:generalupper}}\label{app:generalupper}
For any user $u_i\in U$ with degree $d(u_i)$, the probability that the maximum weight of all the $d(u_i)$ neighboring edges is equal to or less than $v_k$ is given by $(\sum_{i=1}^k p_k)^{d(u_i)}$ where $\sum_{i=1}^k p_k$ is the probability that one edge has weight equal to or less than $v_k$. Thus, the average maximum weight of all $d(u_i)$ edges incident to user $u_i$ is given by $\sum_{k=1}^K v_k((\sum_{i=1}^k p_k)^{d(u_i)}-(\sum_{i=1}^{k-1} p_k)^{d(u_i)})$. Further, the expectation of (\ref{equ:newoptimal}) can be given by
\begin{align}
\mathbb{E}_W[f^\star(G,W)]\leq \frac{1}{2} \sum_{u_i\in U}\sum_{k=1}^K v_k((\sum_{i=1}^k p_i)^{d(u_i)}-(\sum_{i=1}^{k-1} p_i)^{d(u_i)}). \nonumber
\end{align}
The proof is completed.

\section{Proof of Proposition \ref{pro:gridcomplexity}}\label{app:gridcomplexity}

We first note that in $n\times n$ grid, the number of possible chains that start from any given node $u_i$ and have non-decreasing weights is no longer two (toward left or right) as in $1\times n$ grid networks, but still limited to be less than $4K$ (the weight set size $K$ is a given constant). This is because we assume in Algorithm 1, every node will need to use priorities over ties among neighbors whose edge has the same weight and thus the number of possible chains that are relevant to a user's decisions is significantly reduced. Similarly, let $I(u_i)$ be the indicator variable that the length $H(u_i)$ of the longest chain (sequence of edges) that has non-decreasing weights and starts from $u_i$ is greater than $c\log n$ for some constant $c$. Let $I(G)$ be the indicator variable that the linear graph $G$ has at least one such chain with length greater than $c\log n$. Then we have
\begin{align}
\mathbb{E}(I(G))\leq\mathbb{E}(
\sum_{i=1}^n I(u_i))\leq 4K n^2q, 
\end{align}
By using the similar arguments as in linear networks, we can prove that in $n\times n$ grid, the algorithm run in $O(\log n)$ time w.h.p..

\section{Proof of Proposition \ref{pro:gridbound}}\label{app:gridbound}
Based on Proposition \ref{pro:generalupper}, we are able to compute the upper bound of the average total weight under the optimal matching for any given graph. We now consider a $n\times n$ grid and that the weight set $V=\{v_1=1,v_2=1+\Delta\}$ and the weight probability distribution $P=\{p_1=\frac{1}{2},p_2=\frac{1}{2}\}$ are given. Note that, in this gird, there are $(n-2)^2$ nodes with degree $4$, $4n-8$ nodes with degree $3$, and $4$ nodes with degree $2$ in the grid. Moreover, the average maximum weight of a node with degree $d$ is $2-(\frac{1}{2})^d$. Thus, we finally obtain the average total weight under the optimal matching in the grid is upper bounded by $\frac{16+15\Delta}{32}n+o(n)$.

Regarding the average total weight of greedy matching, we first note that for each $2\times n$ sub-grid in step 2, the recursive formula for the sequence $\{a_n\}$ is given by
\begin{align}
a_n=\frac{19+15\Delta}{16}+\frac{1}{4}a_{n-1}+\frac{5}{8}a_{n-2}+\frac{1}{8}a_{n-3}.\nonumber
\end{align}
Based on this, we derive the general formula $a_n=\frac{19+15\Delta}{30}n$ by using asymptotic analysis. 

Then, after the matching in Step 2, users in the second row might be unmatched and are available for the matching in step 3. To compute the probability $p_M$ that
any user $u_{i}$ in the first row is matched to her vertical neighbor $u_{i+n}$ in the second row in step 2, we first define the proposal probability $y_k^r$ ($y_k^l$) that $u_i$ receives a proposal from her right (left) neighbor given the weight between them is $v_k$. Moreover, we have that $y_k^r$ and $y_k^l$ satisfy the following recursive formulas:
\begin{align}
y_k^r=\frac{k}{K}(1-\sum_{t=k+1}^K \frac{y_t^r}{K}),\nonumber\\
y_k^l=\frac{k}{K}(1-\sum_{t=k}^K \frac{y_t^r}{K}).\nonumber
\end{align}

When $K=2$, we can obtain $y_1^r=\frac{1}{4}$, $y_2^r=1$, $y_1^l=\frac{4}{15}$ and $y_2^l=\frac{2}{3}$. Based on this, we can compute the matching probability $p_M=\frac{4}{15}$ for any user in the second row in step 2. After removing these matched users, the remaining graph in the second row becomes a lot of linear segments in step 3 as shown in Fig.~\ref{fig:grid}. We first compute the average number of segments with length $t$ by using probability analysis, which is given by $(1-p_M)^tp_M^2n$. Note that, for a linear segment with size $1\times t$, its greedy matching is denoted by $a_t$. Then, the average weight caused by matched remaining edges in second row is equal to $\sum_{t=1}^{\infty}(1-p_M)^tp_M^2n a_{t}$. 

To compute that, we need the value of $a_t$ for any $t$. We have $a_1=0$, $a_2=\frac{2+\Delta}{2}$, $a_3=\frac{4+3\Delta}{4}$ and $a_t=\frac{4+3\Delta}{4}+\frac{3}{4}a_{t-2}+\frac{1}{4}a_{t-3}$ according to (\ref{equ:greedyrecurrence}). Thus, we can compute the value of $a_t$ for a finite number of $t$, for example, from $t=1$ to $t=100$, and further derive a lower bound for $\sum_{t=2}^{\infty}(1-p_M)^tp_M^2na_{t}$, which is given by $\sum_{t=2}^{100}(1-p_M)^tp_M^2na_{t}\approx (0.288+0.1967\Delta)n$. Note that the value of $\sum_{t=101}^{\infty}(1-p_M)^tp_M^2na_{t}$ is almost zero since $(1-p_M)^t$ decreases exponentially while $a_t$ increases linearly in $t$.

In sum, we can prove that the average total weight under the greedy matching in the $n\times n$ grid is lower bounded by
\begin{align}
\lim_{n\to \infty} \mathbb{E}_W [\hat{f}(G,W)]&\geq(\frac{19+15\Delta}{30}+0.288+0.1967\Delta)\frac{n^2}{2}\nonumber\\
&\geq \frac{0.9213+0.6967\Delta}{2}n^2+o(n^2) \nonumber
\end{align}
where $\mathbb{E}_W [\hat{f}(G,W)]$ denotes the average total weight under the greedy matching in the $n\times n$ grid.

By combining this with the previously derived upper bound for the optimal matching, we finally obtain
\begin{align}
\lim_{n\to \infty} PR(G)\geq \frac{0.9213+0.6967\Delta}{1+0.9375\Delta}.\nonumber
\end{align}
The proof is completed.

\section{Proof of Proposition \ref{pro:performancegnp}}\label{app:performancegnp}

In the random graph $G(n,p)$ with a large user number $n$ and a constant connection probability $p$, the probability that there exist $\sqrt{n}$ nodes that have less edges than $\sqrt{n}$ among them is upper bounded by
\begin{align}
&\lim_{n\to \infty} {n \choose \sqrt{n}}\sum_{i=0}^{\sqrt{n}-1} {\frac{\sqrt{n}(\sqrt{n}-1)}{2} \choose i} y^{i}(1-p)^{\frac{\sqrt{n}(\sqrt{n}-1)}{2}-i}\nonumber\\
&\leq \lim_{n\to \infty} (1-p)^{\frac{n-\sqrt{n}}{2}} {n \choose \sqrt{n}}^2\sum_{i=0}^{\sqrt{n}-1} (\frac{p}{1-p})^i= 0.\nonumber
\end{align}
Therefore, for any $\sqrt{n}$ nodes in $G(n,p)$, there are more than $\sqrt{n}$ edges among them with probability 1.

Let $E_i$ denote the number of edges in the random graph after the greedy matching algorithm adding $i$ edges. The probability that the heaviest edge among $E_i$ edges has weight $v_K$ is $1-(\sum_{k=1}^{K-1}p_k)^{E_i}$. Thus, the probability that the first heaviest edge to add in the greedy algorithm has weight $v_K$ is given by $1-(\sum_{k=1}^{K-1}p_k)^{E_0}$. After adding the first edge in the greedy algorithm, the probability that the heaviest edge to add among the remaining edges has weight $v_K$ is given by $(1-(\sum_{k=1}^{K-1}p_k)^{E_0})(1-(\sum_{k=1}^{K-1}p_k)^{E_1})\geq 1-(\sum_{k=1}^{K-1}p_k)^{E_0}-(\sum_{k=1}^{K-1}p_k)^{E_1}$. Similarly, we have that when $n\to \infty$, the probability that the $(n-\sqrt{n})/2$-th edge to add have weight $v_K$ satisfies
\begin{align}
&\lim_{n\to \infty} 1-\sum_{i=0}^{(n-\sqrt{n})/2-1}(\sum_{k=1}^{K-1}p_k)^{E_i}\nonumber\\
&\geq \lim_{n\to \infty} 1-\frac{n-\sqrt{n}}{2}(\sum_{k=1}^{K-1}p_k)^{E_{(n-\sqrt{n})/2-1}}\nonumber\\
&\geq \lim_{n\to \infty} 1-\frac{n-\sqrt{n}}{2}(\sum_{k=1}^{K-1}p_k)^{\sqrt{n}}= 1,
\end{align}
where the second inequality is because the number of edges $E_{(n-\sqrt{n})/2-1}$ among the remaining $\sqrt{n}+2$ unmatched users is large than $\sqrt{n}$ with probability 1 when $n\to \infty$ as we have proved earlier, 

Therefore, we have that all the first $(n-\sqrt{n})/2$ edges added to the greedy matching have the largest weight $v_K$ with probability 1. Moreover, an obvious upper bound of the total weight under the optimal matching is given by $nv_K/2$ since any two users can be matched with an edge with weight $v_K$ at most. Therefore, the average performance guarantee of the greedy algorithm is given by:
\[\lim_{n\to \infty}\frac{v_K((n-\sqrt{n})/2)}{ v_K n/2}=1.\]
The proof is completed.

\section{Proof of Proposition \ref{pro:complexitygnp}}\label{app:complexitygnp}

Similarly, for any user $u_i$ in the linear network, let $H(u_i)$ denote the length of the longest chain (sequence of edges) that has non-decreasing weights and starts from $u_i$. Let $I(u_i)$ be the indicator variable that $H(u_i)$ is greater than $c\log n$ for some constant $c$. Let $I(G)$ be the indicator variable that the linear graph $G$ has at least one such chain with length greater than $c\log n$. Then we have
\begin{align}
\mathbb{E}(I(G))&\leq\mathbb{E}(
\sum_{i=1}^n I(u_i))=n \mathbb{E}(I(u_1))\nonumber\\
&\leq n(\frac{d(n-1)}{n})^{\log n}q< nd^{c\log n}q, \nonumber 
\end{align}
where $q$ denotes the probability that $c\log n$ consecutive edges has non-decreasing weights. The second inequality is because the expected size of $u_i$'s $c\log n$-th generation is given by $(\frac{d(n-1)}{n})^{\log n}$.

By using the similar arguments as in the proof of Proposition \ref{pro:complexitylinear}, we have $q<n^K (\frac{\max\{p_1,p_2,\dots,p_K\}}{2})^{c\log n}$. Then we obtain:
\[\mathbb{E}(I(G)) \leq (\frac{1}{2})^{-K} n^{K+1}d^{c\log n}(\frac{\max\{p_1,p_2,\dots,p_K\}}{2})^{c\log n}.\]
\[=(\frac{1}{2})^{-K}n^{K+1-c(-\log \frac{\max\{p_1,p_2,\dots,p_K\}}{2}-\log d)}\]
Note that when $d<\frac{2}{\max\{p_1,p_2,\dots,p_K\}}$, there always exists a constant $c>\frac{K+1}{\log2-\log \max\{p_1,p_2,\dots,p_K\}-\log d}$ that makes $\mathbb{E}(I(G))$ converge to $0$ when $n\to \infty$. Therefore, we can conclude that the algorithm will terminate within $c\log n$ iterations w.h.p. according to the first moment method.

\section{Proof of Proposition \ref{pro:gnptree}}\label{app:gnptree}

In $G(n,d/n)$ with $d<1$, to prove \eqref{equ:gnptree}, we first show the connected component of any user $u_i$ has no loops w.h.p., and thus we can analyze the conditional expectation of $\mathbb{E}_{G\sim G(n,d/n)} [x_i(G)| C(u_i)=0]$ assuming the number of loops $C(u_i)$ in $u_i$'s component is zero, instead of directly analyzing $\mathbb{E}_{G\sim G(n,d/n)} [x_i(G)]$ (step 1 below). Then, it remains to show $\mathbb{E}_{G\sim G(n,d/n)} [x_i(G)| C(u_i)=0]$ can be well approximated by $\mathbb{E}_{T\sim T(d),W}[x_{root}(T,W)]$ in the approximated random tree $T(d)$, as both of them are considered in graphs without loops (step 2 below).

\textbf{Step 1: }
In $G(n,d/n)$, there are totally a random number of loops each of which includes at least three users, and for any $t\geq 3$ users, they have totally $\frac{(t-1)!}{2}$ kinds of permutations to form a loop. Thus, the average total number of loops is
\begin{align}
\mathbb{E}(Loop)\hspace{-2pt}=\hspace{-2pt}\sum_{t=3}^n\hspace{-2pt}{n\choose t} \frac{(t\hspace{-2pt}-\hspace{-2pt}1)!}{2}(\frac{d}{n})^t\hspace{-2pt}<\hspace{-2pt}\frac{1}{6}\sum_{t=3}^n (d)^t\hspace{-2pt}<\hspace{-2pt}\frac{(d)^3}{6(1\hspace{-2pt}-\hspace{-2pt}d)},\label{equ:loopnumber}
\end{align}
which implies that regardless of the graph size $n$, there are at most a constant number of loops. Moreover, since \cite{gnpproperty} proves that $G(n,d/n)$ almost surely has no connected components of size larger than $c\log n$ for some constant $c$, the average size of the largest connected component in $G(n,d/n)$ is only $o(n)$. By combining this with \eqref{equ:loopnumber}, the average total number of users in the components with loops should be less than $o(n)\mathbb{E}(Loop)=o(n)$. Therefore, the probability that user $u_i$ is one of these users who are in components with loops is
\begin{align}
\lim_{n\to \infty}Prob(C(u_i)\geq 1)=\frac{o(n)}{n}=0, \label{equ:probnocycle} \end{align}
where $C(u_i)$ denotes the number of loops in the connected component of user $u_i$. Based on \eqref{equ:probnocycle}, we now have:
\begin{align}
&\lim_{n\to \infty} \hspace{-2pt}\mathbb{E}_{G\sim G(n,\frac{d}{n})} [x_i(G)]\hspace{-2pt}=\hspace{-2pt}\lim_{n\to \infty}\hspace{-2pt} \mathbb{E}_{G\sim G(n,\frac{d}{n})} [x_i(G)| C(u_i)\hspace{-2pt}=\hspace{-2pt}0]. \label{equ:firstinequlity}
\end{align}

\textbf{Step 2: }
In this step, our problem becomes to prove that the conditional expectation $\mathbb{E}_{G\sim G(n,d/n)} [x_i(G)| C(u_i)=0]$ can be well approximated by $\mathbb{E}_{T\sim T(d),W}[x_{root}(T,W)]$. Given the connected component of user $u_i$ is a tree (i.e., with $C(u_i)=0$ loop), we start by first showing that the users in the component connect with each other in a similar way as in the random tree $T(d)$. 

We do a breadth-first search (BFS) starting from $u_i$ to explore all its connected users by searching all its direct neighbors prior to moving on to the two-hop neighbors. Note that the number of neighbors we explore from user $u_i$ follows the binomial distribution $B(n-1,d/n)$. In fact, for any user in the BFS tree of $u_i$, the number of new neighbors we directly explore from this user follows the binomial distribution $B(n-m,d/n)$ where $m$ is the number of users that have already been explored currently and cannot be explored again (otherwise a loop would occur). Meanwhile, in $T(d)$, each node gives birth to children according to the same Poisson distribution $Poi(d)$ no matter how large the tree currently is. 

Next, we prove that the difference between the branching under $B(n-m,d/n)$ and $Poi(d)$ is trivial. Actually, for any $t,m \leq c\log n$, the ratio of the probability of getting $t$ from distribution $B(n-m,d/n)$ and the probability of getting $t$ from distribution $Poi(d)$ is bounded by
\begin{align}
1-\frac{1}{\sqrt{n}}\leq \frac {{n-m\choose t}(\frac{d}{n})^t(1-\frac{d}{n})^{n-m-t}}{e^{-d} d^t/(t)!}\leq 1+\frac{1}{\sqrt{n}},\label{equ:ratioboundpoi}
\end{align}
when $n$ is sufficiently large.

We define $\Theta$ as the set of all possible graph structures that the random tree $T(d)$ can have, and $p_1(\theta)$ as the corresponding probability for any structure $\theta\in \Theta$. As the set $\Theta$ includes all possible structures that the BFS tree of $u_i$ can have, we also define $p_2(\theta)$ similarly for the BFS tree. Then, based on \eqref{equ:ratioboundpoi}, for any $\theta\in \Theta$ with user size $s(\theta)< c\log n$, we have
\begin{align}
(1-\frac{1}{\sqrt{n}})^{s(\theta)}\leq\frac{p_2(\theta)}{p_1(\theta)}\leq(1+\frac{1}{\sqrt{n}})^{s(\theta)},\label{equ:sizeratio}
\end{align}
which is because both the probabilities $p_1(\theta)$ and $p_2(\theta)$ are given by the product of all $s(\theta)$ users' individual probability to give birth to a given number of children as in structure $\theta$. 

We now note that the difference between $\mathbb{E}_{T\sim T(d)} [x_{root}(T)]$ and $\mathbb{E}_{G\sim G(n,d/n)} [x_i(G)| C(u_i)=0]$ is determined by the difference between $p_1(\theta)$ and $p_2(\theta)$
\begin{align}
&\mathbb{E}_{T\sim T(d)} [x_{root}(T)]=\sum_{\theta\in\Theta}p_1(\theta) x_{root}(\theta),\label{equ:roottotal}\\
&\mathbb{E}_{G\sim G(n,d/n)} [x_i(G)| C(u_i)=0] =\sum_{\theta\in\Theta}p_2(\theta) x_{root}(\theta).\label{equ:gnptotal}
\end{align}
\eqref{equ:gnptotal} uses $x_{root}(\theta)$ instead of $x_{i}(\theta)$ because the matching weight $x_{i}(\theta)$ of node $u_i$ should be equal to the matching weight $x_{root}(\theta)$ of the root node when the BFS tree of $u_i$ has the same structure $\theta$ as the rooted tree. 

By substituting \eqref{equ:sizeratio} into \eqref{equ:roottotal} and \eqref{equ:gnptotal}, we have
\begin{align}
&\lim_{n\to \infty} |\mathbb{E}_{T\sim T(d)} [x_{root}(T)]-\mathbb{E}_{G\sim G(n,d/n)} [x_i(G)| C(u_i)=0]|\nonumber\\
&=\lim_{n\to \infty} |\sum_{\theta\in\Theta} x_{root}(\theta)(p_1(\theta)-p_2(\theta)) | \nonumber\\
&\leq \lim_{n\to \infty} \frac{v_K}{2}\sum_{\{\theta\in\Theta:s(\theta)\geq c\log n\}} p_1(\theta)+ p_2(\theta) \nonumber\\
&+\frac{v_K}{2}\sum_{\{\theta\in\Theta:s(\theta)<c\log n\}}p_1(\theta)((1+\frac{1}{\sqrt{n}})^{s(\theta)}-1)\nonumber\\
&\overset{\text{(a)}}= \lim_{n\to \infty} \frac{v_K}{2}\frac{1}{\sqrt{n}} \sum_{\{\theta\in\Theta:s(\theta)<c\log n\}} p_1(\theta)s(\theta)\nonumber\\
&\overset{\text{(b)}}\leq \lim_{n\to \infty} \frac{v_K}{2} \frac{1}{\sqrt{n}} \frac{1}{1-d} =0, \label{equ:secondtinequlity}
\end{align}
where we split the analysis in two cases depending on whether the graph structure $\theta$ has size larger than $c\log n$ or not. As we mentioned earlier, \cite{gnpproperty} proves that $G(n,d/n)$ almost surely has no connected components of size larger than $c\log n$ for the constant $c$, thus we have the probability $\sum_{\{\theta\in\Theta:s(\theta)\geq c\log n\}} p_2(\theta)=0$ to derive equality (a). Inequality (b) is due to that the average size of the random tree $T(d)$ is given by $\sum_{\theta\in\Theta} p_1(\theta)s(\theta)=\sum_{t=0}^\infty d^t=\frac{1}{1-d}$ where $d^t$ is the average size of the $t$-th generation as each node gives birth to $d$ children on average. Moreover, according to the first moment method, we also have $\sum_{\{\theta\in\Theta:s(\theta)\geq c\log n\}} p_1(\theta)=0$ for (a). 

Based on \eqref{equ:firstinequlity} and \eqref{equ:secondtinequlity}, we finally prove \eqref{equ:gnptree}.

\section{Proof of Proposition \ref{pro:gnptreeyk}}\label{app:gnptreeyk}

To compute the proposal probability $y_k$, we further define $y_k^c$ to denote the probability that the root node receives a proposal from a child who connects to it with an edge of weight $v_k$ given this child gives birth to $c$ grandchild nodes (happens with probability ${n-1 \choose c}(d/n)^c(1-d/n)^{n-1-c}\to\frac{d^c}{c!}e^{-d}$ as $n\to \infty$). If the edge between the root node and the child has the maximum weight (i.e., $k=K$), the child will send a proposal to the root node only when all $i$ grandchildren that have the maximum weight among the $c$ grandchild nodes (happens with probability ${c \choose i}(p_K)^i (1-p_K)^{c-i}$) either want great-grandchildren (happens with probability $1-y_K$) or have lower priority to be added. Thus, the recursive equation for the proposal probability $y_K^c$ is given as follows:
\[y_K^c\hspace{-2pt}=\hspace{-2pt}\sum_{i=0}^c \hspace{-2pt}{c \choose i}(p_K)^i (1\hspace{-1pt}-\hspace{-1pt}p_K)^{c-i}(\sum_{j=0}^i \hspace{-2pt}{i\choose j}y_K^{i-j}(1\hspace{-1pt}-\hspace{-1pt}y_K)^j\frac{1}{i\hspace{-2pt}-\hspace{-2pt}j\hspace{-2pt}+\hspace{-2pt}1}).\]
Moreover, by considering all possible number of grandchildren that the child can give birth to, we can derive the aggregate recursive equation for the matching possibility $y_K$:
\begin{align}
&y_K=\sum_{c=0}^{\infty} \frac{d^c}{c!}e^{-d} y_K^c
\nonumber\\ 
&=\sum_{c=0}^{\infty} \hspace{-2pt}\frac{d^c}{c!}e^{-d} \sum_{i=0}^c\hspace{-2pt} {c \choose i}(p_K)^i (1\hspace{-2pt}-\hspace{-2pt}p_K)^{c-i}(\sum_{j=0}^i\hspace{-2pt} \frac{{i\choose j}y_K^{i-j}(1\hspace{-2pt}-\hspace{-2pt}y_K)^j}{i-j+1}) \nonumber\\ 
&=e^{-p_K d}\sum_{i=0}^\infty \frac{(p_Kd)^i(1-(1-y_K)^{i+1})}{(i+1)!y_K},\label{equ:first}
\end{align}
after a summation by parts. Note that the term $\frac{1}{y_K}(1-(1-y_K)^{i+1})$ on the right-hand-side of the equation is decreasing in $y_K$ when $y_K\in (0,1)$. Moreover, when $y_K=0$, the RHS of the equation is equal to 1. When $y_K=1$, the RHS of the equation is equal to $\frac{1}{p_Kd}(1-e^{-p_Kd})<1$ for any $d>0$ and $0<p_K<1$. Therefore, there exists a unique solution $y_K^\star$ satisfying the above equation in the interval $(0,1)$.

Then, after the proposal probability $y_{K}$ of the maximum weight has been decided, the proposal probability $y_{K-1}$ now have the highest priority to compute as $w_{K-1}$ becomes the maximum weight among the remaining weights. Similarly, for the proposal probability $y_{K-1}$, we can derive the following recursive equation
\begin{align}
y_{k}=e^{-(p_K+ y_Kp_K) d}\sum_{i=0}^\infty \frac{(p_kd)^i(1-(1-y_k)^{i+1})}{(i+1)!y_k}.\label{equ:second}
\end{align}
Using the similar argument for $y_{K}$, we prove that there exists a unique solution $y_{K-1}^\star$ satisfying the above equation in the interval $(0,1)$ after substituting the solution $y_K^\star$ to (\ref{equ:first}) into (\ref{equ:second}). 

Eventually, for any $k=1,2,\dots K$, we can derive the following recursive equation
\begin{align}
y_{k}=e^{-(p_K+\sum_{j=k+1}^K y_jp_j) d}\sum_{i=0}^\infty \frac{(p_kd)^i(1-(1-y_k)^{i+1})}{(i+1)!y_k}.\nonumber
\end{align}
and prove that there exists a unique solution to the equation.

\section{Proof of Proposition \ref{pro:multiple}}\label{app:multiple}

As shown in Fig~\ref{fig:multiple}, the decomposed sub-graph (1a) has size $1\times n$ and its average total weight $a_n$ under the greedy matching is given by $a_n=\frac{4+3\Delta}{9}+o(n)$ based on \eqref{equ:greedyrecurrence}. 

As for the decomposed sub-graph (1b), only the users with quantity $2$ are left and they form a lot of linear segments. We first compute the average number of segments with length $t$ by using probability analysis, which is given by $\frac{n}{2^{t+2}}$. Note that, for a linear segment with size $1\times t$, its greedy matching is denoted by $a_t$. Then, the average weight caused by matched remaining edges in second row is equal to $\sum_{t=1}^{\infty} \frac{n}{2^{t+2}} a_{t}$. 

To compute that, we need the value of $a_t$ for any $t$. We have $a_1=0$, $a_2=\frac{2+\Delta}{2}$, $a_3=\frac{4+3\Delta}{4}$ and $a_t=\frac{4+3\Delta}{4}+\frac{3}{4}a_{t-2}+\frac{1}{4}a_{t-3}$ according to (\ref{equ:greedyrecurrence}). Thus, we can compute the value of $a_t$ for a finite number of $t$, for example, from $t=1$ to $t=100$ and further derive a lower bound for $\sum_{t=2}^{\infty}\frac{n}{2^{t+2}} a_{t}$, which is given by $\sum_{t=2}^{100}\frac{n}{2^{t+2}} a_{t}\approx (0.16+0.1\Delta)n$. Note that the value of $\sum_{t=101}^{\infty}\frac{n}{2^{t+2}} a_{t}$ is almost zero since $\frac{n}{2^{t+2}}$ decreases exponentially while $a_t$ increases linearly in $t$.

In sum, we can prove that the average total weight under the greedy matching in the linear network of $n$ users is lower bounded by
\begin{align}
\lim_{n\to \infty} \mathbb{E}_W [\hat{f}(G,W)]& \geq (\frac{4+3\Delta}{9}+0.16+0.1\Delta)n\nonumber\\
& \geq (0.604+0.433\Delta)n+o(n) \nonumber
\end{align}
where $\mathbb{E}_W [\hat{f}(G,W)]$ denotes the average total weight under the greedy matching in the linear networks with two possible quantity values 1 and 2.

By combining this with the derived upper bound for the optimal matching using \eqref{equ:newoptimal2}, we finally obtain
\begin{align}
\lim_{n\to \infty}PR(G)\geq \frac{0.604+0.433\Delta}{0.75+0.5\Delta}.\nonumber 
\end{align}
The proof is completed.

\section{Proof of Average Performance Ratio for Non-IID Rounds}\label{app:2rounds}
In a more detailed model, the same user may stay for multiple rounds to fulfill its demand expressed in the first round. This will create dependence between adjacent rounds. To obtain the optimal matching when users stay for multiple rounds, we need to run the matching algorithm that takes into account the entire time horizon assuming full knowledge of the future. More specifically, using full knowledge of arrivals and departures over time, we can construct an enlarged graph where any two nearby users are connected if there is an overlap between their activity intervals. Since users can be matched only once during any activity interval, we can compute the optimal matching using this larger graph.

Here, we further extend our methodology to compare the single-round greedy matching returned by running Algorithm 1 independently in each round with the multi-round optimal matching using future information in the linear matching network model where users stay for $2$ rounds. In this extended linear network model, we assume that there are a population of $n$ users and in the first round each user becomes active in the system with probability $0.5$. Once a user becomes active, it will wait for two rounds to be matched and becomes inactive after. Once inactive, the user joins the system again with the same probability $0.5$ in the following rounds. For illustrative purposes, we also assume the weight set $V=\{v_1=1,v_2=1+\Delta\}$ and uniform weight distribution $p_1=p_2=\frac{1}{2}$ for the sharing benefit between any pair of users. 

Then, to analyze the average performance ratio of the single-round greedy matching as compared to the multi-round optimal matching, we first derive the upper bound of the optimum. Note that, in such a dynamic linear network, each of $n$ users is active for $2/3$ of rounds on average over the entire time horizon. For any user being active, her left/right neighbor is inactive (i.e., unavailable for matching) with probability $1-2/3=1/3$ in her first participating round and keeps being inactive with probability $1/2$ in the second round. Thus, during the two rounds, the user has two available neighbors with probability $25/36$, one available neighbor with probability $10/36$, and no available neighbor with probability $1/36$. Then, by using similar arguments as in Section \ref{sec:generaloptimal}, we can prove that the individual matching weight of this user is upper bounded by 
\[\frac{4+3\Delta}{8}\times\frac{25}{36}+\frac{2+\Delta}{4}\times\frac{10}{36}+0\times\frac{1}{36}=\frac{140+95\Delta}{288}.\]
Hence, the average total weight under the optimal matching in each round is upper bounded by $\frac{140+95\Delta}{864}n$.

Next, we estimate a lower bound of the average total weight under the greedy matching. Note that in each round of the extended linear network, there are $1/3$ of users who turn to active from inactive, $1/3$ of users who keep being inactive, and $1/3$ of users who keep being active since the last round on average. We run the greedy matching algorithm for new participating users independently in each round and these users form a lot of linear segments. Similar to the proof of Proposition \ref{pro:multiple} in Appendix M, we first compute the average number of segments with length $t$ by using probability analysis, which is given by $\frac{4n}{3^{t+2}}$. Note that, for a linear segment with size $1\times t$, its greedy matching is denoted by $a_t$ and we have $a_1=0$, $a_2=\frac{2+\Delta}{2}$, $a_3=\frac{4+3\Delta}{4}$ and $a_t=\frac{4+3\Delta}{4}+\frac{3}{4}a_{t-2}+\frac{1}{4}a_{t-3}$ according to (\ref{equ:greedyrecurrence}). Then, the average total matching weight for all the segments is $\sum_{t=1}^{\infty} \frac{4n}{3^{t+2}} a_{t}$, which is lower bounded by $\sum_{t=1}^{100} \frac{4n}{3^{t+2}} a_{t} \approx0.0816+0.0476\Delta$. Moreover, note that the users in the segments of size $1\times 1$ have no available neighbor in the current round and thus can be left to match with the same kind of users in the next round. Such matching can make an additional matching weight of $\frac{10}{243}+\frac{11}{486}\Delta$. In sum, we can prove that the total weight under the single-round greedy matching is lower bounded by $0.0816+0.0476\Delta+\frac{10}{243}+\frac{11}{486}\Delta=0.1228+0.0703\Delta$.

Finally, by comparing the derived lower bound for the single-round greedy matching with the upper bound for the multi-round optimal matching, we obtain the lower bound of the average performance ratio $PR_t(G)$ for non-IID rounds as follows:
\begin{align}
\lim_{n\to \infty}\lim_{t\to \infty}PR_t(G)\geq \frac{0.1228+0.0703\Delta}{\frac{140+95\Delta}{864}}.\nonumber 
\end{align}
This ratio decreases from $75.8\%$ to $63.9\%$ as weight difference $\Delta$ increases from $0$ to $+\infty$.

The obtained ratio decreases with $\Delta$ and achieves its maximum value when all edges have similar weights (i.e., $\Delta\to 0^+$). This is different from Proposition \ref{pro:guarantee2} for single-round matching in linear networks. We believe this is because the network is sparser in the considered multi-round model where users participate with probability $0.5$. Then, both the single-round greedy matching and the multi-round optimal matching try to match as many pairs as possible, resulting in similar numbers of matched edges and thus a trivial performance/weight gap when all edges have similar weights.

\ifCLASSOPTIONcaptionsoff
  \newpage
\fi

\end{document}